\documentclass[aps,pre,twocolumn,10pt,superscriptaddress,nofootinbib,balancelastpage]{revtex4-2} 
\usepackage[latin1]{inputenc}
\usepackage{amsmath,amssymb}
\usepackage{mathrsfs} 
\usepackage{hyperref}
\usepackage[capitalise]{cleveref}
\usepackage{siunitx}
\usepackage{braket}
\usepackage{calc}
\usepackage{tabularx}
\usepackage{dsfont}
\usepackage[dvipsnames]{xcolor}
\usepackage{ifthen}
\usepackage{graphicx}
\usepackage{pifont}
\usepackage{wasysym}
\usepackage{tikz}
\usetikzlibrary{shapes.geometric}

\allowdisplaybreaks

\newcommand{\orcidicon}{\includegraphics[width=0.32cm]{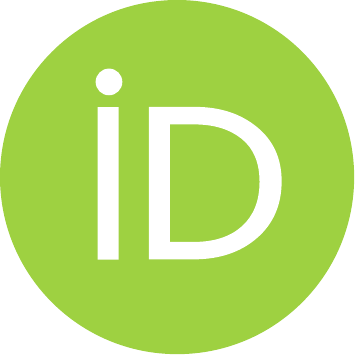}}

\foreach \x in {A, ..., Z}{%
\expandafter\xdef\csname orcid\x\endcsname{\noexpand\href{https://orcid.org/\csname orcidauthor\x\endcsname}{\noexpand\orcidicon}}
}

\newcommand{\Eins}{\mathds{1}}%
\newcommand{\ii}{\mathrm{i}}%
\newcommand{\dif}{\mathrm{d}}%
\newcommand{\Nabla}{\vec{\nabla}}%
\newcommand{\fdif}{\operatorname{\delta}}%
\newcommand{\Fdif}[2]{\frac{\fdif\!#1}{\fdif\!#2}}%
\newcommand{\uu}{\vec{u}}

\newcommand{\norm}[1]{\lVert#1\rVert}%
\newcommand{\rt}{(\vec{r},t)}

\newcommand{\ZT}[1]{\textquotedblleft#1\textquotedblright}%

\newcommand{\FieldTupel}[1]{\mathbf{#1}}
\newcommand{\StateVector}{\FieldTupel{w}}
\newcommand{\StatesVector}{\StateVector}

\newcolumntype{Y}{>{\centering\arraybackslash}X}%
\newcolumntype{Z}{>{\raggedright\arraybackslash}X}%

\newlength{\myl}%
\newcommand{\SUM}[2]{{\setlength{\myl}{\widthof{$\displaystyle\sum_{#1}^{#2}$}*\real{0.5}-\widthof{$\displaystyle\sum$}*\real{0.5}}\sum_{#1}^{#2}\;\hspace{-\the\myl}}}
\newcommand{\INT}[3]{\settowidth{\myl}{$\displaystyle\int_{#1}^{#2}$}{\int_{#1}^{#2}\;\;\;\hspace{-\the\myl}\dif #3}\,}
\newcommand{\TINT}[3]{\settowidth{\myl}{$\int_{#1}^{#2}$}{\int_{#1}^{#2}\!\ifthenelse{\equal{#1#2}{}}{}{\;\;\;\;\hspace{-\the\myl}}\dif #3}\,}%
\newcommand{\EINT}[3]{\settowidth{\myl}{$\int_{#1}^{#2}$}{\int_{#1}^{#2}\;\;\;\,\hspace{-\the\myl}\dif #3}\,}

\newcommand{\EinrueckabstandI}{\quad\,}%
\begin{document}
	
\title{Derivation and analysis of a phase field crystal model for a mixture of active and passive particles}
\author{Michael te Vrugt\orcidA{}}
\affiliation{Institut f\"ur Theoretische Physik, Westf\"alische Wilhelms-Universit\"at M\"unster, 48149 M\"unster, Germany}
\affiliation{Center for Soft Nanoscience (SoN), Westf\"alische Wilhelms-Universit\"at M\"unster, 48149 M\"unster, Germany}

\author{Max Philipp Holl\orcidB{}}
\affiliation{Institut f\"ur Theoretische Physik, Westf\"alische Wilhelms-Universit\"at M\"unster, 48149 M\"unster, Germany}

\author{Aron Koch}
\affiliation{Institut f\"ur Theoretische Physik, Westf\"alische Wilhelms-Universit\"at M\"unster, 48149 M\"unster, Germany}

\author{Raphael Wittkowski\orcidC{}}
\email{raphael.wittkowski@uni-muenster.de}
\affiliation{Institut f\"ur Theoretische Physik, Westf\"alische Wilhelms-Universit\"at M\"unster, 48149 M\"unster, Germany}
\affiliation{Center for Soft Nanoscience (SoN), Westf\"alische Wilhelms-Universit\"at M\"unster, 48149 M\"unster, Germany}
\affiliation{Center for Nonlinear Science (CeNoS), Westf\"alische Wilhelms-Universit\"at M\"unster, 48149 M\"unster, Germany}

\author{Uwe Thiele\orcidD{}}
\email{u.thiele@uni-muenster.de}
\homepage{http://www.uwethiele.de}
\affiliation{Institut f\"ur Theoretische Physik, Westf\"alische Wilhelms-Universit\"at M\"unster, 48149 M\"unster, Germany}
\affiliation{Center for Nonlinear Science (CeNoS), Westf\"alische Wilhelms-Universit\"at M\"unster, 48149 M\"unster, Germany}
\affiliation{Center for Multiscale Theory and Computation (CMTC), Westf\"alische Wilhelms-Universit\"at M\"unster, 48149 M\"unster, Germany}

\begin{abstract}		
We discuss an active phase field crystal (PFC) model that describes a mixture of active and passive particles. First, a microscopic derivation from dynamical density functional theory (DDFT) is presented that includes a systematic treatment of the relevant orientational degrees of freedom. Of particular interest is the construction of the nonlinear and coupling terms. This allows for interesting insights into the microscopic justification of phenomenological constructions used in PFC models for active particles and mixtures, the approximations required for obtaining them, and possible generalizations. Second, the derived model is investigated using linear stability analysis and nonlinear methods. It is found that the model allows for a rich nonlinear behavior with states ranging from steady periodic and localized states to various time-periodic states. The latter include standing, traveling, and modulated waves corresponding to spatially periodic and localized traveling, wiggling, and alternating peak patterns and their combinations.
\end{abstract}
\maketitle

\section{\label{introduction}Introduction}
Active particles such as animals, bacteria, and artificial microswimmers, which transform energy into directed motion, are of significant importance to modern physics both due to their remarkable collective nonequilibrium behavior and their significant technological and biological relevance \cite{ShaebaniWWGR2020,BechingerdLLRVV2016,MarchettiJRLPRS2013,GompperEtAl2020}. Of particular interest in current research are mixtures of active and passive particles, which can exhibit remarkable dynamics. An example is motility-induced phase separation (MIPS) \cite{CatesT2015}, i.e., liquid-gas phase separation in a system of active particles that can occur even for purely repulsive interactions. The state diagram for MIPS has been found to be affected by the presence of passive particles \cite{RodriguezAMRV2020}. Further phenomena exhibited by active-passive mixtures include separation of active and passive particles \cite{StenhammarWMC2015}, Hopf bifurcations leading to moving clusters \cite{WittkowskiSC2017}, collapse in a circular shear cell \cite{HoellLM2019}, transitions from gas-like behavior to oscillations and collapses \cite{SturmerSS2019}, and the existence of different demixing types \cite{AiZZ2020}. This has also led to an increased interest in theoretical models describing such mixtures \cite{WittkowskiSC2017,HoellLM2019,AlaimoV2018,FengH2018,TakatoriB2015b}. Moreover, topological defects in active nematics can be modeled as mixtures of active and passive particles \cite{ShankarRMB2018,ShankarM2019}. The study of active-passive mixtures in polymer systems is of biological relevance as it allows one to describe phase separation in DNA strands \cite{SmrekK2017,SmrekK2018}. Research on such mixtures also has a variety of possible applications, such as controlling the properties of passive materials using active dopants \cite{RamananarivoDP2019,KummelSLVB2015,vanderMeerDF2016,vanderMeerFD2016} and effects of swimming microorganisms such as algae on their environment \cite{JeanneretPKP2016,Visser2007,WilliamsJTP2021}. Finally, whenever the active particles are immersed in a passive solvent -- as is the case in almost all experimental or theoretical scenarios where active matter is considered -- we are, strictly speaking, already dealing with a mixture of active and passive particles. Note that some of the mentioned properties of active-passive mixtures do also occur in \ZT{standard} active matter models without the passive component. For instance, although in active phase field crystal models \cite{MenzelL2013} (see below) the onset of motion of clusters normally occurs via drift pitchfork bifurcations \cite{OphausGT2018,OphausKGT2020b}, Hopf bifurcations can also be responsible \cite{OphausKGT2020}.
 
A useful tool for the description of active matter is provided by field theories. These exist in a large variety of forms, ranging from simple Cahn-Hilliard-type models \cite{SpeckBML2014,SolonSCKT2018}, such as active model~B \cite{WittkowskiTSAMC2014} and its extensions active model~B+ \cite{TjhungNC2018}, active model~H \cite{TiribocchiWMC2015,CatesT2018}, and active model I+ \cite{teVrugtFHHTW2022}, to complex predictive models \cite{BickmannW2019b,BickmannW2020,BickmannBJW2020,BickmannBW2022} and active thin-film models \cite{TrinschekSJT2020,LoisyEL2020}. Of particular importance is \textit{dynamical density functional theory} (DDFT), developed in Refs.\ \cite{Evans1979,Munakata1989,Kawasaki1994,MarconiT1999,MarconiT2000,ArcherE2004,Fraaije1993} and reviewed in Refs.\ \cite{teVrugtLW2020,teVrugtW2022}, which can be derived systematically from the equations of motion of the individual particles and thus constitutes a microscopic description of a system. DDFTs for active systems have been developed in a large variety of forms \cite{WensinkL2008,WittkowskiL2011,PototskyS2012,SharmaB2017,MonchoD2020,WittmannMMB2017,EnculescuS2011,WittmannB2016,MenzelSHL2016,HoellLM2017,HoellLM2018,HoellLM2019,AroldS2020b,BleyDM2021,BleyHDM2021}. Moreover, DDFT has been successfully applied in related fields such as cancer growth \cite{ChauviereLC2012,AlSaediHAW2018} and disease spreading \cite{teVrugtBW2020,teVrugtBW2020b}.

Due to the complexity of DDFT, that normally represent integro-differential, i.e., nonlocal, equations, \textit{phase field crystal} (PFC) models are a popular alternative. Originally proposed on a phenomenological basis \cite{ElderKHG2002,ElderG2004,BerryGE2006}, they were found to be a local approximation to DDFT \cite{ElderPBSG2007,vanTeeffelenBVL2009}. PFC models have been extended into a variety of directions, allowing to describe, e.g., binary mixtures \cite{HuangEP2010,TahaDMEH2019,ElderKHG2002,HollAT2020,RobbinsATK2012,AnkudinovG2022} and particles with orientational degrees of freedom \cite{Loewen2010,WittkowskiLB2010,WittkowskiLB2011,WittkowskiLB2011b}. The active PFC model, an extension of the PFC model to active matter which was proposed and derived from DDFT in Refs.\ \cite{MenzelL2013,MenzelOL2014}, has also been extended, allowing one to describe systems on curved surfaces \cite{PraetoriusVWL2018}, self-spinning particles \cite{HuangML2020}, particles with inertia \cite{AroldS2020,teVrugtJW2021}, particles with nonreciprocal interactions \cite{HollAGKOT2021}, and mixtures of active and passive particles \cite{AlaimoV2018,HollAGKOT2021} -- an extension suggested already in the first article on active PFC models \cite{MenzelL2013}. Further investigations of active PFC models can be found in Refs.\ \cite{AlaimoPV2016,ChervanyovGT2016,OphausGT2018,OphausKGT2020,OphausKGT2020b,HuangLV2022}. A review of PFC models is given by Refs.\ \cite{EmmerichEtAl2012,StarodumovAN2022}, the derivation of PFC models from DDFT is discussed in Refs.\ \cite{ArcherRRS2019,teVrugtLW2020}. Reference \cite{HollAGKOT2021} discusses the occurrence of localized states in a number of passive and active PFC models.

When considering PFC models for mixtures, it is interesting that microscopic derivations \cite{HuangEP2010,TahaDMEH2019} give relatively complicated coupling terms, whereas models proposed on a phenomenological basis typically employ a very simple coupling \cite{ElderKHG2002,HollAT2020,RobbinsATK2012,HollAGKOT2021}. A similar observation can be made for active PFC models: The derivation of the active PFC model from DDFT in Ref.\ \cite{MenzelOL2014} is focused on the dynamical part, whereas the free energy was simply imported from phenomenological models \cite{MenzelL2013}. 
Microscopic treatments, on the other hand, show that the free energy in a PFC model with orientational dynamics can be very complex \cite{Loewen2010,WittkowskiLB2010,WittkowskiLB2011,WittkowskiLB2011b}. Consequently, a systematic microscopic derivation of an active PFC model for a binary mixture is required to investigate both what the general form looks like and which assumptions are required to recover simple phenomenological models, thereby providing insights into their range of applicability and into ways to improve them.

The aim of the present work is twofold: First, we derive a general PFC model for a mixture of active and passive particles from a microscopic DDFT and obtain from it a simpler minimal model (which we refer to as \textit{model 1}), thereby revealing both the necessary approximations and possible generalizations. Second, we investigate the active binary PFC model using linear and nonlinear methods. It is found to have interesting properties that differ from those of simple active PFC models and passive models for binary mixtures.

This article is structured as follows: The minimal binary active PFC model is introduced in \cref{equations}. A general PFC model is derived from DDFT in \cref{derivation}. Approximations leading to the minimal model are discussed in \cref{derivation2}. In \cref{linear}, we perform a linear stability analysis of the PFC model. Nonlinear results are shown in \cref{nonlinear}. We conclude in \cref{conc}.

\section{\label{equations}Governing equations}
We start by introducing the central model of this work based on phenomenological arguments. In the active PFC model \cite{MenzelL2013,MenzelOL2014,OphausGT2018,OphausKGT2020}, a system is described using a conserved scalar field $\psi\rt$ (depending on position $\vec{r}$ and time $t$), which describes the dimensionless deviation from a mean particle number density, and a nonconserved vector field $\vec{P}\rt$, which measures the polarization. They follow the dynamic equations 
\begin{align}
\partial_t \psi &= \Nabla^2\frac{\delta F}{\delta \psi} - v_0 \Nabla \cdot \vec{P},\label{psi}\\ 
\partial_t \vec{P}&= \Nabla^2\frac{\delta F}{\delta \vec{P}} - D_r \frac{\delta F}{\delta \vec{P}} - v_0 \Nabla\psi\label{p}.
\end{align}
Here, $v_0$ is the activity of the particles and $F$ is a free energy functional. This functional can be written as
\begin{equation}
F = F_\mathrm{pfc} + F_\mathrm{P},
\label{freeenergy}
\end{equation}
where 
\begin{equation}
F_\mathrm{pfc} = \INT{}{}{^dr}\bigg(\frac{1}{2}\psi(\epsilon_\psi+(q_\psi^2+\Nabla^2)^2)\psi + \frac{1}{4}(\psi + \bar{\psi})^4 \bigg) 
\label{freeenergypfc}
\end{equation}
is a Swift-Hohenberg-type free energy \cite{SwiftH1977,EmmerichEtAl2012} in $d$ spatial dimensions and
\begin{equation}
F_\mathrm{P}=\INT{}{}{^dr}\bigg(\frac{C_1}{2}\vec{P}^2 + \frac{C_2}{4}\vec{P}^4\bigg)    
\label{freeenergypol}
\end{equation}
is the orientational contribution to $F$. For $C_1 <0$ and $C_2>0$, the system exhibits spontaneous polarization. Most treatments only consider the case $C_1 > 0$ and $C_2 = 0$ where no spontaneous polarization occurs \cite{OphausGT2018,MenzelOL2014}. In addition to the parameters $C_1$ and $C_2$, the free energy depends on the parameters $\epsilon_\psi$ (scaled shifted temperature), $q_\psi^2$ (critical wavenumber), and $\bar{\psi}$ (dimensionless mean density). We mark these parameters with a subscript $\psi$ since they can differ for the components of a mixture, for example if its components have different freezing temperatures \cite{HollAT2020}.

For the case $v_0 = 0$, \cref{psi} reduces to the standard PFC model for passive systems \cite{EmmerichEtAl2012,ThieleARGK2013}. It has the form of a gradient dynamics, where the system evolves toward the minimum of a free energy. The passive PFC model shares this property with DDFT \cite{teVrugtLW2020} and many other soft matter models \cite{Cahn1965,PototskyBMT2005,Doi2011,XuTQ2015,ThieleAP2016}. In the active case, however, the coupling terms $-v_0 \Nabla \cdot \vec{P}$ and $-v_0 \Nabla\psi$ in \cref{psi,p} break this structure since they are nonvariational, i.e., they cannot be obtained from a free energy. This is a typical feature of active matter models \cite{WittkowskiTSAMC2014}.

The active PFC model given by \cref{psi,p} is a two-field model for a single species of active particles. The scalar field $\psi$ and vector field $\vec{P}$ describe the density and orientational order (polarization) of the single particle species of a one-component system. A PFC model can, however, also describe mixtures of different particle species employing a larger number of fields. For a simple case of a binary mixture of passive particles, see Refs.\ \cite{TahaDMEH2019,HollAT2020}. In contrast, here, we consider a binary mixture of passive particles with density $\phi$ and active particles with density $\psi$ and polarization $\vec{P}$. In this case, we have to add to \cref{psi,p} the equation
\begin{equation}
\partial_t \phi = \Nabla^2\frac{\delta F}{\delta \phi}.\label{phi}  
\end{equation}
Note that no modification of \cref{psi} is required, since the coupling between the different particle species is purely variational (for a case of nonvariational coupling, see Section~4.2 of Ref.\ \cite{HollAGKOT2021}). The variational coupling is introduced by adding a mixed term $F_{\mathrm{coup}}[\psi,\phi]$ to the free energy~\eqref{freeenergy}, which now takes the form
\begin{equation}
F = F_\mathrm{pfc}[\psi] + F_\mathrm{pfc}[\phi] + F_\mathrm{P}[\vec{P}] + F_{\mathrm{coup}}[\psi,\phi].
\label{freeenergycomplete}
\end{equation}
It is common \cite{ElderKHG2002,HollAT2020,RobbinsATK2012} to use the simplest nontrivial form
\begin{equation}
F_{\mathrm{coup}} = \INT{}{}{^dr}a\psi\phi,  
\label{freeenergycoup}
\end{equation}
where $a$ is a coupling constant. Inserting \cref{freeenergycomplete,freeenergypfc,freeenergypol,freeenergycoup} into \cref{psi,p,phi}, setting $C_2=0$, and carrying out the functional derivatives gives the equations of motion
\begin{align}
\begin{split}
\partial_t \psi &= \Nabla^2((\epsilon_\psi + (q_\psi^2+\Nabla^2)^2)\psi + (\bar{\psi}+\psi)^3) \\
&\quad\:\! -v_0 \Nabla\cdot \vec{P} + a\Nabla^2\phi,\label{uno}
\end{split}\\
\partial_t \vec{P} &= (\Nabla^2 - D_r)C_1 \vec{P} - v_0 \Nabla \psi\label{tres},\\
\begin{split}
\partial_t \phi &= \Nabla^2((\epsilon_\phi + (q_\phi^2+\Nabla^2)^2)\phi + (\bar{\phi}+\phi)^3) \\
&\quad\:\! +a\Nabla^2\psi,\label{dos}
\end{split}
\end{align}
where by convention we have explicitly introduced the mean densities $\bar{\psi}$ and $\bar{\phi}$ as parameters, i.e., $\INT{}{}{^dr}\psi=0 $ and $\INT{}{}{^dr}\phi=0 $.
From here on, we refer to the model given by \cref{uno,dos,tres} as \ZT{model 1}. It is closely related to active binary PFC models used in previous works \cite{HollAGKOT2021,AlaimoV2018}, although there are small differences (Ref.\ \cite{HollAGKOT2021} uses $\epsilon_\psi=\epsilon_\phi$, $C_2 \neq 0$, and $q_\psi=q_\phi=1$, Ref.\ \cite{AlaimoV2018} uses a nonlinear instead of a linear coupling term).

\section{\label{derivation}Derivation of general model}
After having introduced model 1 on a phenomenological basis, we now aim for a systematic derivation from the microscopic dynamics of the particles. In doing so, we will make use of a variety of previous results on the derivation of PFC models from DDFT. The \ZT{standard} case of a one-component system with no orientational degrees of freedom is discussed in Refs.\ \cite{vanTeeffelenBVL2009,ArcherRRS2019,EmmerichEtAl2012}. Binary systems were considered in Refs.\ \cite{HuangEP2010,TahaDMEH2019}, orientational degrees of freedom in Refs.\  \cite{Loewen2010,AchimWL2011,WittkowskiLB2010,WittkowskiLB2011,WittkowskiLB2011b}, and active particles in Refs.\ \cite{MenzelL2013,MenzelOL2014}.

The starting point is the DDFT
\begin{align}
\partial_t \rho_\psi(\vec{r},\uu,t) &= \Nabla \cdot D_T \cdot \bigg( \beta\rho_\psi(\vec{r},\uu,t) \Nabla \frac{\delta F}{\delta \rho_\psi(\vec{r},\uu,t)}\bigg) \nonumber\\
&\quad\:\! + D_R\partial_\varphi\bigg(\beta\rho_\psi(\vec{r},\uu,t) \partial_\varphi \frac{\delta F}{\delta \rho_\psi(\vec{r},\uu,t)}\bigg) \nonumber\\
&\quad\:\! - \Nabla\cdot D_T\cdot\bigg( \rho_\psi(\vec{r},\uu,t)\frac{v}{D_\parallel}\uu\bigg),
\label{firsteq}\\
\partial_t \rho_\phi\rt &= D \Nabla \cdot\bigg(\beta\rho\rt\Nabla\frac{\delta F}{\delta \rho_\phi \rt}\bigg)\label{secondeq},
\end{align}
where $\rho_\psi$ and $\rho_\phi$ are ensemble-averaged one-body densities (this is why no noise terms are required \cite{ArcherR2004,teVrugtLW2020,teVrugt2020}), $\beta$ is the thermodynamic beta, $D_T = D_\parallel \uu\otimes\uu + D_\perp(\Eins -\uu \otimes \uu$) with constants $D_\parallel$ and $D_\perp$, dyadic product $\otimes$, and unit matrix $\Eins$ is the translational diffusion tensor of the (active) field $\rho_\psi$, $D_R$ its rotational diffusion constant, $D$ the translational diffusion constant of the (passive) field $\rho_\phi$, $\uu(\varphi) = (\cos(\varphi),\sin(\varphi))^\mathrm{T}$ with polar angle $\varphi$ the orientation, $F$ an equilibrium free energy functional, and $v$ the active propulsion speed. The density $\rho_\psi$ gives the probability of finding a particle of species $\psi$ with orientation $\uu$ at position $\vec{r}$ at time $t$, whereas $\rho_\phi$ gives the probability of finding a particle of species $\phi$ (assumed to have no orientational degrees of freedom) at position $\vec{r}$ at time $t$. Equation \eqref{firsteq} is a standard active DDFT for uniaxial particles \cite{WensinkL2008} in two spatial dimensions and identical to the one employed in Ref.\ \cite{MenzelOL2014}. The more general case of active particles with arbitrary shape in three spatial dimensions was considered in Ref.\ \cite{WittkowskiL2011}. Equation \eqref{secondeq} is a standard DDFT for passive particles. As discussed in Refs.\ \cite{MarconiT1999,ArcherE2004,WittkowskiL2011}, DDFT can be derived systematically from the microscopic Langevin equations for active and passive particles using the \ZT{adiabatic approximation}, where it is assumed that the pair correlations in the nonequilibrium system are equal to those of an equilibrium system with the same one-body density. This assumption, which is required to close the equations of motion for the one-body density, is not exactly true, but it is a good approximation for many systems of interest \cite{teVrugtLW2020}.

To derive the corresponding PFC model, we need to approximate both the dynamical equations and the free energy. Although these approximations are, as shown by \citet{ArcherRRS2019}, intimately connected, it is helpful to consider them separately. We start with the free energy. In (D)DFT, it is (ignoring external potentials) given by
\begin{equation}
F = F_{\mathrm{id}} + F_{\mathrm{exc}}.
\label{freeenergyddft}
\end{equation}
Here, $F_{\mathrm{id}}$ is the ideal gas free energy. It is known exactly and given by\footnote{We do not write the dependence on time $t$ in the equations for $F$ to emphasize that we are dealing with equilibrium functionals.}
\begin{equation}
\begin{split}
F_{\mathrm{id}} &= \beta ^{-1}\INT{}{}{^2r}\INT{0}{2\pi}{\varphi}\rho_\psi(\vec{r},\uu)(\ln(\Lambda_\psi^2\rho_\psi(\vec{r},\uu)) -1) \\
&\quad\:\! +\beta^{-1}\INT{}{}{^2r}\rho_\phi(\vec{r})(\ln(\Lambda_\phi^2\rho_\phi(\vec{r})) -1),
\label{idealgas}
\end{split}
\end{equation}
where $\Lambda_\psi$ and $\Lambda_\phi$ are the (irrelevant) thermal de Broglie wavelengths, which may differ for the two species.
 
The second term in \cref{freeenergyddft} is the excess free energy $F_\mathrm{exc}$. It is not known exactly and has to be approximated. A systematic way of doing this is a functional Taylor expansion \cite{Evans1979,WittkowskiLB2011,EmmerichEtAl2012,teVrugtLW2020} around a homogeneous state up to fourth order, which gives
\begin{widetext}
\begin{align}
\beta F_\mathrm{exc} &= - \frac{1}{2}\INT{}{}{^2r}\INT{}{}{^2r_1}\INT{0}{2\pi}{\varphi}\INT{0}{2\pi}{\varphi_1}c_{\psi\psi}(\vec{r},\vec{r}_1,\uu,\uu_1)\Delta\rho_\psi(\vec{r},\uu)\Delta\rho_\psi(\vec{r}_1,\uu_1) \notag\\
&\quad\:\! - \INT{}{}{^2r}\INT{}{}{^2r_1}\INT{0}{2\pi}{\varphi}c_{\psi\phi}(\vec{r},\vec{r}_1,\uu)\Delta\rho_\psi(\vec{r},\uu)\Delta\rho_\phi(\vec{r}_1)- \frac{1}{2}\INT{}{}{^2r}\INT{}{}{^2r_1}c_{\phi\phi}(\vec{r},\vec{r}_1)\Delta\rho_\phi(\vec{r})\Delta\rho_\phi(\vec{r}_1) \notag\\
&\quad\:\! - \frac{1}{6}\INT{}{}{^2r}\INT{}{}{^2r_1}\INT{}{}{^2r_2}\INT{0}{2\pi}{\varphi}\INT{0}{2\pi}{\varphi_1}\INT{0}{2\pi}{\varphi_2}c_{\psi\psi\psi}(\vec{r},\vec{r}_1,\vec{r}_2,\uu,\uu_1,\uu_2)\Delta\rho_\psi(\vec{r},\uu)\Delta\rho_\psi(\vec{r}_1,\uu_1)\Delta\rho_\psi(\vec{r}_2,\uu_2) \notag\\
&\quad\:\! - \frac{1}{2}\INT{}{}{^2r}\INT{}{}{^2r_1}\INT{}{}{^2r_2}\INT{0}{2\pi}{\varphi}\INT{0}{2\pi}{\varphi_1}c_{\psi\psi\phi}(\vec{r},\vec{r}_1,\vec{r}_2,\uu,\uu_1)\Delta\rho_\psi(\vec{r},\uu)\Delta\rho_\psi(\vec{r}_1,\uu_1)\Delta\rho_\phi(\vec{r}_2) \notag\\
&\quad\:\! - \frac{1}{2}\INT{}{}{^2r}\INT{}{}{^2r_1}\INT{}{}{^2r_2}\INT{0}{2\pi}{\varphi}c_{\psi\phi\phi}(\vec{r},\vec{r}_1,\vec{r}_2,\uu)\Delta\rho_\psi(\vec{r},\uu)\Delta\rho_\phi(\vec{r}_1)\Delta\rho_\phi(\vec{r}_2) \notag\\
&\quad\:\! - \frac{1}{6}\INT{}{}{^2r}\INT{}{}{^2r_1}\INT{}{}{^2r_2}c_{\phi\phi\phi}(\vec{r},\vec{r}_1,\vec{r}_2)\Delta\rho_\phi(\vec{r})\Delta\rho_\phi(\vec{r}_1)\Delta\rho_\phi(\vec{r}_2) \notag\\
&\quad\:\! - \frac{1}{24}\INT{}{}{^2r}\INT{}{}{^2r_1}\INT{}{}{^2r_2}\INT{}{}{^2r_3}\INT{0}{2\pi}{\varphi}\INT{0}{2\pi}{\varphi_1}\INT{0}{2\pi}{\varphi_2}\INT{0}{2\pi}{\varphi_3}c_{\psi\psi\psi\psi}(\vec{r},\vec{r}_1,\vec{r}_2,\vec{r}_3,\uu,\uu_1,\uu_2,\uu_3)\label{functionaltaylorexpansion}\\
&\qquad\qquad \Delta\rho_\psi(\vec{r},\uu)\Delta\rho_\psi(\vec{r}_1,\uu_1)\Delta\rho_\psi(\vec{r}_2,\uu_2)\Delta\rho_\psi(\vec{r}_3,\uu_3) \notag\\
&\quad\:\! - \frac{1}{6}\INT{}{}{^2r}\INT{}{}{^2r_1}\INT{}{}{^2r_2}\INT{}{}{^2r_3}\INT{0}{2\pi}{\varphi}\INT{0}{2\pi}{\varphi_1}\INT{0}{2\pi}{\varphi_2}c_{\psi\psi\psi\phi}(\vec{r},\vec{r}_1,\vec{r}_2,\vec{r}_3,\uu,\uu_1,\uu_2) \notag\\
&\qquad\qquad\:\!\Delta\rho_\psi(\vec{r},\uu)\Delta\rho_\psi(\vec{r}_1,\uu_1)\Delta\rho_\psi(\vec{r}_2,\uu_2)\Delta\rho_\phi(\vec{r}_3) \notag\\
&\quad\:\! - \frac{1}{4}\INT{}{}{^2r}\INT{}{}{^2r_1}\INT{}{}{^2r_2}\INT{}{}{^2r_3}\INT{0}{2\pi}{\varphi}\INT{0}{2\pi}{\varphi_1}c_{\psi\psi\phi\phi}(\vec{r},\vec{r}_1,\vec{r}_2,\vec{r}_3,\uu,\uu_1)\Delta\rho_\psi(\vec{r},\uu)\Delta\rho_\psi(\vec{r}_1,\uu_1)\Delta\rho_\phi(\vec{r}_2)\Delta\rho_\phi(\vec{r}_3) \notag\\
&\quad\:\! - \frac{1}{6}\INT{}{}{^2r}\INT{}{}{^2r_1}\INT{}{}{^2r_2}\INT{}{}{^2r_3}\INT{0}{2\pi}{\varphi}c_{\psi\phi\phi\phi}(\vec{r},\vec{r}_1,\vec{r}_2,\vec{r}_3,\uu)\Delta\rho_\psi(\vec{r},\uu)\Delta\rho_\phi(\vec{r}_1)\Delta\rho_\phi(\vec{r}_2)\Delta\rho_\phi(\vec{r}_3) \notag\\
&\quad\:\! - \frac{1}{24}\INT{}{}{^2r}\INT{}{}{^2r_1}\INT{}{}{^2r_2}\INT{}{}{^2r_3}c_{\phi\phi\phi\phi}(\vec{r},\vec{r}_1,\vec{r}_2,\vec{r}_3)\Delta\rho_\phi(\vec{r})\Delta\rho_\phi(\vec{r}_1)\Delta\rho_\phi(\vec{r}_2)\Delta\rho_\phi(\vec{r}_3). \notag
\end{align}
\end{widetext}
We have used the direct correlation functions \cite{EmmerichEtAl2012}
\begin{equation}
c_{\varrho_1,\dotsc,\varrho_n}=\beta\frac{\delta^n F}{\delta \rho_{\varrho_1} \dotsb \delta \rho_{\varrho_n}} \bigg|_{\rho_{\varrho_i}=\bar{\rho}_{\varrho_i}}
\label{dcf}
\end{equation}
with $\varrho_i \in \{\psi,\phi\}$, as well as the deviations
\begin{equation}
\Delta \rho_{\varrho_i} = \rho_{\varrho_i} - \bar{\rho}_{\varrho_i} 
\end{equation}
from the homogeneous reference densities $\bar{\rho}_{\varrho_i}$. Irrelevant zeroth-order and first-order contributions have been ignored. (Zeroth-order contributions have no effect at all, first-order contributions shift the chemical potential by a constant \cite{ArcherRTK2012}, but vanish after applying the gradient operator.)

Next, we have to define the fields $\psi$, $\vec{P}$, and $\phi$ in terms of the microscopic densities $\rho_\psi$ and $\rho_\phi$. In principle, there are two possibilities for a physical interpretation of the order parameters in a binary PFC model: The first option is to introduce a shifted rescaled total density field $n_1$ and a rescaled density difference field $n_2$, defined as $n_1 = (\rho_\phi + \rho_\psi - \bar{\rho})/\bar{\rho}$ and $n_2 = (\rho_\psi - \rho_\phi)/\bar{\rho}$, where $\bar{\rho} = \bar{\rho}_\psi + \bar{\rho}_\phi$ is the total homogeneous reference density \cite{ElderPBSG2007}. In this case, $n_1$ describes the structure and $n_2$ the composition of the particle distribution \cite{EmmerichEtAl2012}. Here, however, we choose the second option \cite{TahaDMEH2019} of letting $\psi$ and $\phi$ describe the shifted rescaled densities of the two different particle species, i.e., the different fields in the theory directly correspond to the two different particle types. This has the advantage that we can clearly distinguish between an active field $\psi$ and a passive field $\phi$. The existence of these two options shows the importance of providing a microscopic derivation for the binary PFC model, since otherwise the physical interpretation of the fields and therefore the implications of theoretical results for experiments are not clear.

For the field $\rho_\psi$, we use the Cartesian orientational expansion \cite{teVrugtW2020}
\begin{equation}
\rho_\psi(\vec{r},\uu) = \bar{\rho}_\psi(1+\psi(\vec{r}) + \vec{P}(\vec{r})\cdot\uu + \dotsb),   
\label{psiexpansion}
\end{equation}
where
\begin{equation}
\psi(\vec{r}) = \frac{1}{2\pi\bar{\rho}_\psi}\bigg(\INT{0}{2\pi}{\varphi}\rho_\psi(\vec{r},\uu)\bigg)-1
\label{psidefinition}
\end{equation}
is the rescaled shifted density of the active particles and
\begin{equation}
\vec{P}(\vec{r})=\frac{1}{\pi\bar{\rho}_\psi}\INT{0}{2\pi}{\varphi}\rho_\psi(\vec{r},\uu)\uu   
\end{equation}
is their local polarization. The expansion \eqref{psiexpansion} is equivalent to a Fourier series \cite{teVrugtW2020,teVrugtW2019c}, which is widely used in derivations of field theories for active particles \cite{WittkowskiSC2017,BickmannW2019b,BickmannW2020}. By truncating the expansion \eqref{psiexpansion} after the polarization term $\vec{P}$, we are neglecting the contribution of the nematic order to the free energy (as is common in active PFC models). More general expressions including the nematic order can be found in Refs.\ \cite{WittkowskiLB2010,WittkowskiLB2011,WittkowskiLB2011b}. For $\rho_\phi$, where an orientational expansion is not required since this field has no orientational dependence, we simply define
\begin{equation}
\phi(\vec{r}) = \frac{\rho_\phi(\vec{r})}{\bar{\rho}_\phi} -1.
\label{phidefinition}
\end{equation}
 
The next step is the expansion of the ideal gas free energy. If we insert \cref{psiexpansion,phidefinition} into \cref{idealgas} and expand the result up to fourth order (to allow for stable crystals) in both fields, we find
\begin{equation}
\begin{split}
\beta F_\mathrm{id} &= \INT{}{}{^2r}  
\bigg(2\pi\bar{\rho}_\psi\bigg(\frac{\psi^2}{2}- \frac{\psi^3}{6} + \frac{\psi^4}{12} + \frac{\vec{P}^2}{4}\\ 
&\quad\:\! - \frac{\psi \vec{P}^2}{4} + \frac{\psi^2\vec{P}^2}{4}+ \frac{\vec{P}^4}{32}\bigg) 
+\bar{\rho}_\phi\bigg(\frac{\phi^2}{2} - \frac{\phi^3}{6}+\frac{\phi^4}{12}\bigg)\!\bigg),
\end{split}    
\label{idealgasfreeenergy}\raisetag{4em}%
\end{equation}
where we have dropped irrelevant terms of zeroth and first order.

The excess free energy is more difficult to treat. We follow Ref.\ \cite{WittkowskiLB2011b} and start with a Fourier expansion of the direct correlation functions. Our treatment of symmetries parallels that of Refs.\ \cite{WittkowskiLB2010,WittkowskiLB2011,WittkowskiLB2011b,WittkowskiSC2017,BickmannW2020,BickmannW2019b,BickmannBJW2020,teVrugtFHHTW2022}, with differences arising from the fact that not all particles have an intrinsic orientation. Let us consider the direct correlation functions $c_{\varrho_1,\dotsc,\varrho_n}(\{\vec{r}_i\},\{\varphi_i\})$, where $\varrho_1,\dotsc,\varrho_j = \psi$ and $\varrho_{j+1},\dotsc,\varrho_n = \phi$. Translational and rotational invariance implies that they can be parametrized as $c_{\varrho_1,\dotsc,\varrho_n}(\{R_i \},\{\tilde{\varphi}_{R_i}\},\{\tilde{\varphi}_i\})$ with
\begin{align}
\vec{r}-\vec{r}_i &= R_i \uu(\varphi_{R_i}),\\
\tilde{\varphi}_{R_i} &= 
\begin{cases}
\varphi - \varphi_{R_i} &\text{for }j \geq 1,\\
\varphi_{R_1}- \varphi_{R_{i+1}} &\text{for }j=0,
\end{cases}\label{varphirtilde}\\
\tilde{\varphi}_i &= \varphi - \varphi_i.\label{varphitilde}
\end{align}
Note that there is no dependence on $\tilde{\varphi}_i$ for $j=0$ (if $\varrho_i = \phi$ for all $i$ in \cref{dcf}, then no functional derivatives with respect to functions depending on $\varphi$ are taken on the right-hand side, such that the direct correlation function cannot depend on $\varphi$) or $j=1$ (since then there is no angle $\varphi_1$ and thus \cref{varphitilde} is undefined). Moreover, translational and rotational invariance implies that $c_{\phi\phi}$ also does not depend on $\tilde{\varphi}_{R_i}$ (there is no angle $\varphi_{R_2}$ in this case, such that, since $j=0$, \cref{varphirtilde} is undefined).

The Fourier expansion of the correlation function then reads
\begin{align}
&c_{\varrho_1,\dotsc,\varrho_n}(\{R_i \},\{\tilde{\varphi}_{R_i}\},\{\tilde{\varphi}_i\})
\nonumber\\ &= \sum_{m_1,\dotsc,m_{n-1} = -\infty}^{\infty}\sum_{m'_1,\dotsc,m'_{j-1} = -\infty}^{\infty}c^{\underline{m},\underline{m}'}_{\varrho_1,\dotsc,\varrho_n}(\{R_i \}) 
\label{fourierexpansion}\\
\nonumber&\quad\; e^{\ii (\underline{m}\cdot\underline{\tilde{\varphi}}_R + \underline{m}'\cdot \underline{\tilde{\varphi}})}
\end{align}
with the coefficients
\begin{align}
&c^{\underline{m},\underline{m}'}_{\varrho_1,\dotsc,\varrho_n}(\{R_i \}) 
\nonumber\\&= \frac{1}{(2\pi)^{n+j-2}}\INT{0}{2\pi}{\underline{\tilde{\varphi}}_R}\INT{0}{2\pi}{\underline{\tilde{\varphi}}} c_{\varrho_1,\dotsc,\varrho_n}(\{R_i \},\{\tilde{\varphi}_{R_i}\},\{\tilde{\varphi}_i\})
\nonumber\\&\quad\; e^{ - \ii (\underline{m}\cdot\underline{\tilde{\varphi}}_R + \underline{m}'\cdot \underline{\tilde{\varphi}})},
\label{fouriercoefficients}
\end{align}
the imaginary unit $\ii$, and the vectors $\underline{\tilde{\varphi}}_R = (\tilde{\varphi}_{R_1},\dotsc,\tilde{\varphi}_{R_{n-1-\delta_{j0}}})$, $\underline{\tilde{\varphi}} = (\tilde{\varphi}_1,\dotsc,\tilde{\varphi}_{j-1})$, $\underline{m} = (m_1, \dotsc, m_{n-1-\delta_{j0}})$, and $\underline{m}' = (m_1',\dotsc, m_{j-1}')$.

For our purposes, it is sufficient to consider the terms with $m_i = 0$ and $m'_i = 0, \pm 1$ in \cref{fourierexpansion}. The reason for why we only need $m'_i = 0, \pm 1$ is straightforward: We only consider orientational order parameters of zeroth and first order, which is equivalent to considering only the zeroth- and first-order contributions in a Fourier expansion of the one-body density \cite{teVrugtW2020}. Due to the orthogonality of the basis functions in a Fourier expansion, higher-order contributions of the direct correlation functions will then vanish after performing the angular integrals in \cref{functionaltaylorexpansion}.

We now replace $\Delta\rho_\psi \to \bar{\rho}_\psi(\psi + \vec{P}\cdot\uu )$ and $\Delta \rho_\phi \to \bar{\rho}_\phi \phi$ in \cref{functionaltaylorexpansion}. If we also insert \cref{fourierexpansion} into \cref{functionaltaylorexpansion} and perform the integrals over $\varphi,\varphi_1,\dotsc$, every term will have a prefactor $c^{\underline{m},\underline{m}'}_{\varrho_1,\dotsc,\varrho_n}(\{R_i \}) e^{-\ii\underline{m}\cdot\underline{\varphi}_R}$. The next step (which is helpful to illustrate why we can restrict ourselves to $m_i = 0$) is a gradient expansion, which is used to obtain a local free energy. The general procedure is explained in \cref{gradientexpansion}.  A gradient expansion corresponds to a Taylor expansion of the spatial Fourier transformations of these prefactors. This expansion is truncated at zeroth order, except for terms resulting from the second-order terms in \cref{functionaltaylorexpansion} that do not involve $\vec{P}$ (we assume gradients of the polarization to be small compared to density gradients), which are truncated at fourth order. It is common in the literature to gradient expand the second-order contribution in the functional Taylor expansion up to fourth \cite{EmmerichEtAl2012} and the third- and fourth-order contributions up to zeroth order \cite{ArcherRRS2019}, and it is common in active PFC models to truncate the gradient expansion at zeroth order for the polarization terms \cite{MenzelL2013}. If we have $m_i \neq 0$ for at least one $i$, the zero-wavelength component of the expansion coefficients is
\begin{align}
& \frac{1}{(2\pi)^{n-1}}\INT{0}{\infty}{\underline{R}}\INT{0}{2\pi}{\underline{\varphi_R}}R_1\dotsb R_{n-1} c^{\underline{m},\underline{m}'}_{\varrho_1,\dotsc,\varrho_n}(\{R_i \})
\\\nonumber&  e^{\ii (\underline{m}\cdot\underline{\tilde{\varphi}}_R+\underline{m}'\cdot\underline{\tilde{\varphi}})}e^{-\ii \underline{\vec{k}}\cdot \underline{\vec{r}}}\Big\rvert_{\underline{\vec{k}}=\underline{\vec{0}}}= 0,
\end{align}
where $\underline{R} = (R_1,\dotsc, R_{n-1})$, $\underline{\varphi_R} = (\varphi_{R_1},\dotsc, \varphi_{R_{n-1}})$, $\underline{\vec{k}} = (\vec{k}_1,\dotsc, \vec{k}_{n-1})$, and $\underline{\vec{r}} = (R_1 \uu(\varphi_{R_1}),\dotsc,R_{n-1} \uu(\varphi_{R_{n-1}}))$ are vectors (with the wavenumbers $\vec{k}_i$). The only terms for which nonzero wavenumbers are considered are contributions of the second order in the functional Taylor expansion \eqref{functionaltaylorexpansion} (i.e., from $c_{\psi\psi}$, $c_{\psi\phi}$, and $c_{\phi\phi}$) that do not involve $\vec{P}$. The contributions resulting from $c_{\psi\psi}$ and $c_{\psi\phi}$ vanish when the integral over $\varphi$ is performed if any $m_i\neq 0$, and $c_{\phi\phi}$ has no angular dependence anyway. Consequently, we can restrict ourselves to $\underline{m}=\underline{0}$. All superscripts of the expansion coefficients of the direct correlation function from now on are entries of $\underline{m}'$.

Combining the Fourier and the gradient expansion and performing the angular integrals, we find
\begin{align}
F_\mathrm{exc} &= - \INT{}{}{^2r} \bigg(\frac{1}{2} (A_1 \psi^2 + A_2 \psi \Nabla^2 \psi + A_3 \psi \Nabla^4\psi
\nonumber\\&\quad\, + A_4 \vec{P}^2 + A_5 \psi\phi + A_6 \psi \Nabla^2\phi
\nonumber\\&\quad\, + A_7 \psi \Nabla^4 \phi + A_8 \phi^2 + A_9\phi\Nabla^2\phi 
\nonumber\\&\quad\, + A_{10}\phi\Nabla^4\phi) + \frac{1}{3}(A_{11}\psi^3  + A_{12} \psi\vec{P}^2
\nonumber\\&\quad\, + A_{13}\psi^2\phi + A_{14}\vec{P}^2\phi + A_{15}\psi\phi^2 
\label{localform}\\&\quad\, + A_{16}\phi^3) + \frac{1}{4}(A_{17} \psi^4  + A_{18}\psi^2\vec{P}^2
\nonumber\\&\quad\, + A_{19}\vec{P}^4 + A_{20}\psi^3 \phi + A_{21}\psi\vec{P}^2\phi
\nonumber\\&\quad\,  + A_{22}\psi^2\phi^2 + A_{23}\vec{P}^2\phi^2 
\nonumber\\&\quad\, + A_{24}\psi\phi^3 + A_{25}\phi^4)\bigg).
\nonumber
\end{align}
The expansion coefficients $A_1$--$A_{25}$ are defined in \cref{coefficients}. They are defined with a prefactor $1/n$ for terms of $n$-th order in the fields. Defining also $B_1 = 2\pi \beta^{-1}\bar{\rho}_\psi$ and $B_2 = \beta^{-1}\bar{\rho}_\phi$, we find the complete free energy
\begin{align}
F &= \INT{}{}{^2r}\bigg( B_1\bigg(\frac{\psi^2}{2}-\frac{\psi^3}{6} + \frac{\psi^4}{12} \notag
\\&\quad\:\! + \frac{\vec{P}^2}{4} - \frac{\psi \vec{P}^2}{4} + \frac{\psi^2\vec{P}^2}{4}+ \frac{\vec{P}^4}{32}\bigg) \notag
\\&\quad\:\! +B_2\bigg(\frac{\phi^2}{2} - \frac{\phi^3}{6}+\frac{\phi^4}{12}\bigg) \notag
\\&\quad\:\! -\bigg(\frac{1}{2} (A_1 \psi^2 + A_2 \psi \Nabla^2 \psi + A_3 \psi \Nabla^4\psi \notag
\\&\quad\:\! + A_4 \vec{P}^2 + A_5 \psi\phi + A_6 \psi \Nabla^2\phi \notag
\\&\quad\:\!+ A_7 \psi \Nabla^4 \phi + A_8 \phi^2 + A_9\phi\Nabla^2\phi \label{largefreeenergy}
\\&\quad\:\!+ A_{10}\phi\Nabla^4\phi) + \frac{1}{3}(A_{11}\psi^3  + A_{12} \psi\vec{P}^2 \notag
\\&\quad\:\! + A_{13}\psi^2\phi + A_{14}\vec{P}^2\phi + A_{15}\psi\phi^2 \notag
\\&\quad\:\! + A_{16}\phi^3) + \frac{1}{4}(A_{17} \psi^4  + A_{18}\psi^2\vec{P}^2 \notag
\\&\quad\:\! + A_{19}\vec{P}^4 + A_{20}\psi^3 \phi + A_{21}\psi\vec{P}^2\phi \notag
\\&\quad\:\!  + A_{22}\psi^2\phi^2 + A_{23}\vec{P}^2\phi^2 \notag
\\&\quad\:\! + A_{24}\psi\phi^3 + A_{25}\phi^4)\bigg)\!\bigg). \notag
\end{align}
Next, we wish to rewrite the free energy \eqref{largefreeenergy} in a more familiar form. In particular, we want to eliminate terms proportional to $\psi^3$ and $\phi^3$. For this purpose, we introduce the rescaled fields
\begin{align}
\tilde{\psi} &=\psi - \Delta\psi,\\
\tilde{\phi} &=\phi - \Delta\phi
\end{align}
with the shifts
\begin{widetext}
\begin{align}
&\Delta\psi=\frac{-12A_{16}A_{20}+48A_{11}A_{25}+24A_{25}B_1-16A_{11}B_2-6A_{20}B_2-8B_1B_2}{9 A_{20}A_{24} - 144 A_{17}A_{25} + 48 A_{25}B_1 + 48 A_{17}B_2 - 16 B_1 B_2}, \label{deltapsi}\\
&\Delta\phi =\frac{-12A_{11}A_{24}+48A_{16}A_{17}+24A_{17}B_2-16A_{16}B_1-6A_{24}B_1 -8B_1B_2}{9 A_{20}A_{24} - 144 A_{17}A_{25} + 48 A_{25}B_1 + 48 A_{17}B_2 - 16 B_1 B_2}\label{deltaphi}.
\end{align}
\end{widetext}
Note that \cref{deltapsi,deltaphi} have their complicated form only due to the presence of third- and fourth-order contributions in the functional Taylor expansion \eqref{functionaltaylorexpansion}. Otherwise, we would simply have
\begin{equation}
\Delta \psi = \Delta \phi =\frac{1}{2}.   
\end{equation}
The general free energy \eqref{largefreeenergy} can then be written in the form
\begin{align}
\nonumber F &= \INT{}{}{^2r}\bigg( \frac{1}{2}\tilde{\psi}(\tilde{\epsilon}_\psi + \tilde{A}_3(\tilde{q}_\psi^2 + \Nabla^2)^2)\tilde{\psi} + \tilde{A}_{17} \frac{\tilde{\psi}^4}{4}\\
\nonumber& \quad\:\!+ \frac{\tilde{A}_4}{2}\vec{P}^2 + \frac{\tilde{A}_{19}}{4}\vec{P}^4\\
\nonumber&\quad\:\!+\frac{1}{2}\tilde{\phi}(\tilde{\epsilon}_\phi + \tilde{A}_{10}(\tilde{q}_\phi^2 + \Nabla^2)^2)\tilde{\phi} + \tilde{A}_{25} \frac{\tilde{\phi}^4}{4}\\
&\quad\:\!+ \tilde{\psi}(\tilde{\epsilon}_{\mathrm{coup}} + \tilde{A}_7(\tilde{q}_{\mathrm{coup}}^2 + \Nabla^2)^2)\tilde{\phi}\label{largefreeenergy1}\\
\nonumber&\quad\:\!- \frac{1}{3}(\tilde{A}_{12} \tilde{\psi}\vec{P}^2+ \tilde{A}_{13}\tilde{\psi}^2\tilde{\phi} + \tilde{A}_{14}\vec{P}^2\tilde{\phi}  + \tilde{A}_{15}\tilde{\psi}\tilde{\phi}^2) 
\\\nonumber&\quad\:\! - \frac{1}{4}(\tilde{A}_{18}\tilde{\psi}^2\vec{P}^2 + \tilde{A}_{20}\tilde{\psi}^3 \tilde{\phi} + \tilde{A}_{21}\tilde{\psi}\vec{P}^2\tilde{\phi}
\\\nonumber&\quad\:\!  + \tilde{A}_{22}\tilde{\psi}^2\tilde{\phi}^2 + \tilde{A}_{23}\vec{P}^2\tilde{\phi}^2 + \tilde{A}_{24}\tilde{\psi}\tilde{\phi}^3 )\bigg), 
\end{align}
with the scaled shifted temperatures $\tilde{\epsilon}_i$, the wavenumbers $\tilde{q}_i$ (with $i=\psi,\phi,\mathrm{coup}$), and the rescaled expansion coefficients $\tilde{A}_3$-$\tilde{A}_{25}$. All coefficients are listed in \cref{coefficients2}. Note that the coefficients $A_{20}$-$A_{24}$ are not affected by the shifts (such that, e.g., $\tilde{A}_{20}=A_{20}$). Terms of zeroth and first order in fields that arise from the redefinition of the fields have been ignored as they do not contribute to the dynamics. We have also slightly changed the ordering of the terms compared to \cref{largefreeenergy}. The advantage of using the form \eqref{largefreeenergy1} is that we can see some structure: The first terms up to $\tilde{A}_{25}\tilde{\phi}^4/4$ are the familiar contributions from model 1 (with an additional contribution $\tilde{A}_{19}\vec{P}^4/4$) with a free energy that is an even polynomial in $\tilde{\psi}$, $\tilde{\phi}$, and $\vec{P}$. Then, there are terms coupling $\tilde{\psi}$ and $\tilde{\phi}$ at linear order in the equation of motion, thereby including higher-order spatial gradients. We have written them in such a way that they have the same (Swift-Hohenberg-like) structure as the other linear terms (but without a prefactor $1/2$, as this prefactor would not vanish after the functional derivatives). Note that $\tilde{\epsilon}_{\mathrm{coup}}$, unlike $\tilde{\epsilon}_\psi$ and $\tilde{\epsilon}_\phi$, has no ideal gas contribution (see \cref{epspsi,epsphi,tildeepsiloncoup}) and can therefore, strictly speaking, not be interpreted as a scaled shifted temperature. Finally, we have terms that lead to nonlinear couplings between the fields in the equations of motion. These arise from the ideal gas term (coupling between $\tilde{\psi}$ and $\vec{P}$) and from higher-order contributions in \cref{functionaltaylorexpansion} (coupling of all fields).

Next, we turn to the dynamical part. For \cref{secondeq}, this is rather simple: We insert the parametrization \eqref{phidefinition} into \cref{secondeq} and make a constant mobility approximation, i.e., we replace the prefactor $D\beta(1+\phi)$ by $D\beta$ to arrive at
\begin{equation}
\partial_t \phi = \frac{D\beta}{\bar{\rho}_\phi}\Nabla^2\frac{\delta F}{\delta \phi},
\label{phidim}
\end{equation}
which is identical to \cref{phi} up to rescaling. The situation is more complex for \cref{firsteq}, where we can follow Refs.\ \cite{WittkowskiLB2011b,MenzelOL2014}. We insert the parametrization \eqref{psiexpansion}, and use the fact that \cite{WittkowskiLB2011b}
\begin{equation}
\frac{\delta F}{\delta \rho(\vec{r},\uu)} = \frac{1}{2\pi\bar{\rho}_\psi}\bigg(\frac{\delta F}{\delta \psi(\vec{r})} + 2\frac{\delta F}{\delta \vec{P}(\vec{r})}\cdot\uu\bigg).
\end{equation}
Moreover, we make (as done in Ref.\ \cite{MenzelOL2014}) the approximation $D_\parallel\approx D_\perp = D_0$, such that $D_T = D_0 \Eins$. Finally, we make the constant mobility approximations $D_i\beta (1+\psi+\vec{P}\cdot\uu) \approx D_i\beta$ (with $i=T,R$) for the passive terms in \cref{firsteq}. Then, \cref{firsteq} yields
\begin{align}
\nonumber \partial_t (\psi + \vec{P}\cdot\uu)
&= \frac{D_0\beta}{2\pi\bar{\rho}_\psi}  \Nabla^2\frac{\delta F}{\delta \psi}+ \frac{D_0\beta}{\pi\bar{\rho}_\psi}\Nabla^2\frac{\delta F}{\delta \vec{P}}\cdot \uu
\\&\quad\,+ \frac{D_R\beta}{2\pi\bar{\rho}_\psi} \partial_\varphi^2 \frac{\delta F}{\delta \psi} + \frac{D_R\beta}{\pi\bar{\rho}_\psi} \partial_\varphi^2 \frac{\delta F}{\delta \vec{P}}\cdot\uu \label{firsteqapprox}
\\\nonumber &\quad\,-\Nabla\cdot(\psi + \vec{P}\cdot\uu)v\uu.
\end{align}
Integrating \cref{firsteqapprox} over $\varphi$ and using $\partial_\varphi^2 (\delta F/\delta \psi) = 0$, $\partial_\varphi^2 \uu = - \uu$, and $\TINT{0}{2\pi}{\varphi}\uu = \vec{0}$ gives
\begin{equation}
\partial_t \psi=  \frac{D_0\beta}{2\pi\bar{\rho}_\psi} \Nabla^2\frac{\delta F}{\delta \psi} - \frac{v}{2}\Nabla \cdot \vec{P}.\label{psidim}
\end{equation}
Similarly, multiplying \cref{firsteqapprox} by $\uu$ and integrating over $\varphi$ using $\TINT{0}{2\pi}{\varphi}\uu\otimes\uu= \pi\Eins$ gives
\begin{equation}
\partial_t \vec{P}= \frac{D_0\beta}{\pi\bar{\rho}_\psi} \Nabla^2\frac{\delta F}{\delta \vec{P}} - \frac{D_R\beta}{\pi\bar{\rho}_\psi}\frac{\delta F}{\delta \vec{P}} - v\Nabla \psi\label{pdim}.
\end{equation}
Equations \eqref{psidim} and \eqref{phidim} are equivalent to Eqs.\ (6) and (7) in Ref.\ \cite{MenzelOL2014} and, up to rescaling, identical to \cref{psi,p} in the present work. Note that the coupling term $v_0\Nabla\cdot(\psi \vec{P})$ employed in Refs.\ \cite{AlaimoPV2016,AlaimoV2018} does not appear in the derivation presented here.

What is left to do now is nondimensionalization. If we write $t = t_0 \tilde{t}$, $\vec{r} = r_0 \tilde{\vec{r}}$, $\tilde{\psi}= \psi_0 \underline{\tilde{\psi}}$, $\vec{P} = P_0 \tilde{\vec{P}}$, $\tilde{\phi} = \phi_0 \underline{\tilde{\phi}}$, and $F = \beta^{-1} \tilde{F}$ with dimensionless time $\tilde{t}$, dimensionless position $\tilde{\vec{r}}$, dimensionless free energy $\tilde{F}$, rescaled fields $\underline{\tilde{\psi}}$, $\tilde{\vec{P}}$, and $\underline{\tilde{\phi}}$, and constants
\begin{align}
r_0 &=\sqrt[6]{\frac{\beta\tilde{A}_3^2}{\tilde{A}_{17}}},\\
t_0 &= \frac{2\pi \tilde{A}_3\bar{\rho}_\psi}{D_0 \tilde{A}_{17}},\\
\psi_0 &= \sqrt[6]{\frac{1}{\beta^2\tilde{A}_3\tilde{A}_{17}}},\\
P_0 &= \sqrt{2}\psi_0,\\
\phi_0 &=\sqrt[12]{\frac{\tilde{A}_{17}}{\beta^4\tilde{A}_3^2\tilde{A}_{25}^3}},
\end{align}
and drop all tildes and underlines, we obtain the free energy
\begin{align}
\nonumber F &= \INT{}{}{^2r}\bigg(\frac{1}{2}\psi(\epsilon_\psi + (q_\psi^2 + \Nabla^2)^2)\psi + \frac{1}{4}\psi^4\\ 
\nonumber &\quad\,+\frac{1}{2}\phi(\epsilon_\phi + G_\phi(q_\phi^2 + \Nabla^2)^2)\phi + \frac{1}{4}\phi^4 \\
\nonumber &\quad\,+  \psi(\epsilon_{\mathrm{coup}} + G_{\mathrm{coup}}(q_{\mathrm{coup}}^2 + \Nabla^2)^2)\phi\\
&\quad\,+ \frac{1}{2}C_1 \vec{P}^2 + \frac{1}{4}C_2 \vec{P}^4 + C_3 \psi \vec{P}^2 \label{freeenergymodel3a}\\
\nonumber &\quad\,+ C_4 \psi^2 \vec{P}^2  + C_5 \vec{P}^2\phi + C_6 \vec{P}^2\phi^2\\
\nonumber &\quad\,+C_7 \psi \vec{P}^2\phi +a_2 \psi^2\phi + a_3 \psi \phi^2\\
\nonumber &\quad\,+ a_4 \psi^3\phi + a_5 \psi^2\phi^2 + a_6 \psi\phi^3\bigg)
\end{align}
and the dynamic equations
\begin{align}
\partial_t \psi &= \Nabla^2 \frac{\delta F}{\delta \psi} - v_0 \Nabla \cdot \vec{P},\label{psi4}\\
\partial_t \vec{P} &= \Nabla^2\frac{\delta F}{\delta \vec{P}} - D_r\frac{\delta F}{\delta \vec{P}} - v_0 \Nabla \psi\label{p4},\\ 
\partial_t \phi &= M_\phi \Nabla^2 \frac{\delta F}{\delta \phi}\label{phi4}.
\end{align}
Definitions of the nondimensionalized coefficients $\epsilon_i$, $q_i$, $G_i$, $C_1$-$C_7$, $a_2$-$a_6$, $v_0$, and $M_\phi$ are listed in \cref{rescaling}. Equations \eqref{freeenergymodel3a}--\eqref{phi4} constitute the most general PFC model considered in this work, which will be referred to as \textit{model 3a}. Essentially, the rescaling gives us five free parameters ($r_0$, $t_0$, $\psi_0$, $P_0$, and $\phi_0$) that we can use to set five coefficients to one. We have chosen the prefactors of $\psi^4/4$, $\phi^4/4$, and $\Nabla^4\psi^2/2$ in the free energy and the mobilities of $\psi$ and $\vec{P}$. Thereby, our equations resemble as closely as possible those used in previous work \cite{HollAGKOT2021}. We have absorbed the prefactors $1/n$ into the rescaled coefficients (except in cases where the standard passive PFC model \cite{EmmerichEtAl2012} also contains them).

When comparing model~3a to model~1 (\cref{uno,dos,tres}), we can note that model~3a only contains $\psi$ and $\phi$, whereas \cref{uno,dos,tres} additionally contain parameters $\bar{\psi}$ and $\bar{\phi}$ measuring the total particle number. This is a matter of notation and not a physical difference. The replacements $\psi \to \bar{\psi} + \psi$ and $\phi \to \bar{\phi} + \phi$, which are required to obtain \cref{uno,dos,tres}, can, mathematically, be made without any problem since they simply correspond to a redefinition of the fields $\psi$ and $\phi$ that (since we drop terms up to first order) only affects the nonlinear terms in \cref{freeenergymodel3a} (of which, in model 1, only $\psi^4/4$ and $\phi^4/4$ are left). Physically, however, there is a lot to unpack here. Often, it is argued that the parameter $\bar{\psi}$ is a measure for the total particle number of the species $\psi$ \cite{OphausGT2018,OphausKGT2020}. This, however, is in need of further justification if the parameter $\bar{\psi}$ simply arises from a redefinition of $\psi$ that, by itself, does not necessarily have any physical meaning. 

The key to understanding this aspect is the physical interpretation of the constants $\bar{\rho}_\psi$ and $\bar{\rho}_\phi$ in \cref{psidefinition,phidefinition}. There are two options, which, for simplicity, we discuss using \cref{phidefinition}. First, we can choose for $\bar{\rho}_\phi$ the homogeneous density of the liquid state. In this case, the total number of the particles of type $\phi$ is measured by the parameter $\bar{\rho}_\phi$, whereas the field $\phi$ only measures their spatial distribution. The advantage of this approach is that the density of the homogeneous liquid state is certainly the most natural choice for $\bar{\rho}_\phi$. However, the disadvantage is that, in this case, a change of the total particle number affects almost all parameters of the PFC model. In many cases, it is desirable to have only \textit{one} parameter that corresponds to the total particle number, since this facilitates to study the effects of varying it by means of, e.g., numerical continuation \cite{HollAT2020}. What we can make use of here is that $\bar{\rho}_\phi$ does not have to be the actual liquid density \cite{EmmerichEtAl2012}. In principle -- and this is the second option -- we can make any choice that is convenient (although we should ensure that it does not deviate too much from the actual density to ensure that the functional Taylor expansion \eqref{functionaltaylorexpansion} can be truncated). It has also been argued \cite{ArcherRTK2012} that one can use $\bar{\rho}_\phi$ to eliminate the $\phi^3$ term from the PFC free energy functional. Namely, if one writes the free energy as 
\begin{equation}
F = \INT{}{}{^2r}(f(\rho_\phi)+ \textrm{gradient terms})
\end{equation}
with a local free energy density $f$, we can Taylor expand $f$ up to fourth order around $\rho_\phi=\bar{\rho}_\phi$ and then choose $\bar{\rho}_\phi$ in such a way that the third derivative $f^{(3)}(\bar{\rho}_\phi)$ vanishes. This procedure, however, is less general than the rescaling of $\phi$ employed here since it is not guaranteed that there is a density $\bar{\rho}_\phi$ for which  $f^{(3)}(\bar{\rho}_\phi)=0$. For example, in the case of an ideal gas, we have  $f^{(3)}(\rho_\phi)= -\beta^{-1}/\rho_\phi^2$, which does not vanish regardless of the choice of $\bar{\rho}_\phi$.

If we use $\bar{\rho}_\psi$ and $\bar{\rho}_\phi$ as adjustable parameters, we could in principle use \cref{mphi} to set $M_\phi=1$, namely by choosing them in such a way that
\begin{equation}
\frac{\bar{\rho}_\phi}{\bar{\rho}_\psi}=\frac{2\pi D}{D_0}\sqrt{\frac{\tilde{A}_{25}}{\tilde{A}_{17}}}.\label{mphi2}
\end{equation}
We will, however, consider the more general case $M_\phi\neq 1$ first (as one generally has to if one uses the physical densities), and only later set $M_\phi=1$ to obtain model 1 introduced in \cref{equations}.

Once we have fixed $\bar{\rho}_\phi$, the total particle number is controlled by $\TINT{}{}{^2r} \phi$. In particular, we can substitute $\phi \to \bar{\phi} + \phi$, where $\bar{\phi}$ is chosen in such a way that $\TINT{}{}{^2r}\phi = 0$. In this case, the parameter $\bar{\phi}$ is a measure for the total particle number. The same considerations hold for the field $\psi$. Further rescalings of the fields $\psi$ and $\phi$ that have been made during the derivation to simplify the free energy functional do not affect these aspects in principle, although they are, of course, relevant for the quantitative link between the parameters $\bar{\psi}$ and $\bar{\phi}$ and the total particle numbers.

\section{\label{derivation2}Approximations and model hierarchy}
In \cref{equations}, we have introduced the minimal model~1 on phenomenological grounds, whereas the microscopic derivation in \cref{derivation} has led to the very general model~3a. In this section, we will introduce, step by step, the approximations that are required to obtain model 1. Thereby, we can get insights into their physical significance. Moreover, we obtain a hierarchy also containing intermediate models which are more general than model 1, but less general than model 3a. These arise if only some, but not all approximations are made. For ease of notation, we will use $\psi$ and $\phi$ instead of $\bar{\psi} + \psi$ and $\bar{\phi} + \phi$.
 
When comparing the general free energy of model 3a given by \cref{freeenergymodel3a} with the minimal model 1 introduced in \cref{equations}, we can note that model 3a has six properties not present in model 1:
\begin{itemize}
\item \textbf{A}:  Nonlinear terms coupling $\psi$ and $\vec{P}$.
\item \textbf{B}: A nonlinear term proportional to $\vec{P}^4$.
\item \textbf{C}: Nonlinearities coupling $\psi$ and $\phi$ (or $\vec{P}$ and $\phi$).
\item \textbf{D}: Gradient terms in the linear coupling of $\psi$ and $\phi$.
\item \textbf{E}: A factor $G_\phi\neq1$ in front of the term $\frac{1}{2}\phi(q_\phi + \Nabla^2)^2\phi$.
\item \textbf{F}: A mobility $M_\phi$ of the field $\phi$.
\end{itemize}

\begin{table*}
\begin{tabular}{|c|c|c|c|}\hline
\textbf{Level of approximation} & \textbf{With $\boldsymbol{\psi}$-$\boldsymbol{\vec{P}}$ coupling (a)} &\textbf{With $\boldsymbol{\vec{P}^4}$ term (b)} & \textbf{Without $\boldsymbol{\vec{P}^4}$ term (c)} \\ \hline
With nonlinear interaction terms (3) & Model 3a (\cref{freeenergymodel3a}) & Model 3b (\cref{freeenergymodel3b}) & Model 3c (\cref{freeenergymodel3c})\\ \hline
Ramakrishnan-Yussouff approximation (2) & Model 2a (\cref{freeenergymodel2a}) & Model 2b (\cref{freeenergymodel2b}) & Model 2c (\cref{freeenergymodel2c})\\ \hline
Coupling without gradients (1) & Model 1a (\cref{freeenergymodel1a}) & Model 1b (\cref{freeenergymodel1b}) & Model 1c (\cref{freeenergymodel1c})\\ \hline
\end{tabular}
\caption{\label{overview}Overview over the various models.}
\end{table*}  

Apart from the last one, these all affect the free energy. We will now discuss the approximations that remove these features and thereby lead from model 3a to model 1, thereby also classifying intermediate models. In general, models with properties A and B will be denoted by \ZT{a}, models with property B (but not A) by \ZT{b}, models without properties A and B by \ZT{c}, models with properties C and D by \ZT{3}, models with property D but without property C by \ZT{2} and models without properties C and D by \ZT{1}. For example, a model with nonlinear coupling but without nonlinearities in $\vec{P}$ would be \ZT{model 3c}. This naming scheme is illustrated in Table \ref{overview}. Combinations that do not fit into this classification (such as a model that has a nonlinear coupling of $\psi$ and $\phi$ but no gradients in the linear coupling) will not be considered, since their derivation would require keeping some nonstandard third- and fourth-order terms while dropping some standard second-order ones. Notably, properties A-F all concern the passive part of the transport equations and not the nonvariational active term. Therefore, our discussion is also relevant for passive PFC models and will be made with a particular emphasis on the correct passive limit. This gives our results additional significance for applications in materials science (for example in models for passive liquid crystals doped with spherical particles \cite{KopvcanskyEtAl2008,RuhwandlT1997}). Properties E and F do not give rise to additional terms, such that we will not use them to specify an additional class of models (apart from the fact that we reserve the name \ZT{model 1} for the case $G_\phi=M_\phi=1$). 

First, we consider the step from models of type 3 to models of type 2. For this purpose, we make the \textit{Ramakrishnan-Yussouff approximation} \cite{RamakrishnanY1979}, which is very common in the derivation of PFC models from (D)DFT and typically made right from the beginning \cite{EmmerichEtAl2012}. In this approximation, the functional Taylor expansion \eqref{functionaltaylorexpansion} is already truncated at second order. In other words, the expansion coefficients $A_{11}$-$A_{25}$ are all set to zero. The general free energy \eqref{freeenergymodel3a} then simplifies to
\begin{align}
\nonumber F &= \INT{}{}{^2r}\bigg(\frac{1}{2}\psi(\epsilon_\psi + (q_\psi^2 + \Nabla^2)^2)\psi + \frac{1}{4}\psi^4\\ 
\nonumber &\quad\,+\frac{1}{2}\phi(\epsilon_\phi + G_\phi(q_\phi^2 + \Nabla^2)^2)\phi + \frac{1}{4}\phi^4 \\
&\quad\,+ \psi(\epsilon_{\mathrm{coup}} + G_{\mathrm{coup}}(q_{\mathrm{coup}}^2 + \Nabla^2)^2)\phi \label{freeenergymodel2a}\\
\nonumber &\quad\,+ \frac{1}{2}C_1 \vec{P}^2 + \frac{1}{4}C_2 \vec{P}^4 + C_3 \psi \vec{P}^2 + C_4 \psi^2 \vec{P}^2\bigg).
\end{align}
Moreover, as can be seen from the microscopic definitions of the coefficients listed in \cref{coefficients,coefficients2,rescaling}, the microscopic expressions for the expansion coefficients simplify drastically. For example, the coefficients $C_2$-$C_{4}$ now only arise from the ideal gas free energy. The free energy \eqref{freeenergymodel2a} constitutes \textit{model 2a}. It is a model of type 2 by the classification introduced above as it has property D but not property C, and it is a model of type a as it has properties A and B. The main differences to model 3a are
\begin{itemize}
\item There is no nonlinear coupling between the fields $\psi$ and $\phi$ (previously encoded in the coefficients $a_2$-$a_6$ of model 3a). 
\item There is no coupling between $\vec{P}$ and $\phi$.
\end{itemize}
This means that both corresponding aspects of model 3a only arise from higher-order contributions in the functional Taylor expansion, and can thus not be obtained within the Ramakrishnan-Yussouff approximation that is usually considered. This shows three advantages of our approach: First, by considering these higher-order terms, we were able to show that there \textit{can} be a direct coupling of $\vec{P}$ and $\phi$ in a binary mixture of active and passive particles if $\phi$ is the passive field, namely through higher-order effects. Second, we show that models with coupling of $\vec{P}$ and $\phi$ but without nonlinear coupling of $\psi$ and $\phi$ are not reasonable since these couplings have the same physical origin (higher-order terms in \cref{functionaltaylorexpansion}). Third and most importantly, the assumption made in model 1 that fields describing two different particle species couple only linearly (through quadratic terms in the free energy) is quite robust, since nonlinear terms only arise if we go beyond the approximations PFC models are usually based on. This also justifies the phenomenological coupling terms used in simpler models such as the one from Ref.\ \cite{HollAT2020}.

We now move further from models of type 2 to models of type 1. For this purpose, we have to get rid of the higher-order terms in the gradient expansion. To recover model 1, we have to set $A_9$ and $A_{10}$ to zero, while keeping all other terms that are of second order in the functional Taylor expansion \eqref{functionaltaylorexpansion}. A \ZT{naive} argument for dropping these two coefficients while keeping all the others can be found from an expansion in a smallness parameter. Let us assume that the fields are slowly varying in space, such that terms of order $n$ in gradients are of the order $\varepsilon_\mathrm{gr}^n$, where $\varepsilon_\mathrm{gr}$ is a small dimensionless parameter. Moreover, we assume that the ratio of the various correlation functions is given by
\begin{equation}
\frac{c_{\psi\phi}}{c_{\psi\psi}} \approx \frac{c_{\psi\phi}}{c_{\phi\phi}} = \mathcal{O}(\varepsilon_\mathrm{cor}),
\label{scaling}
\end{equation}
where $\varepsilon_\mathrm{cor}$ is another small dimensionless parameter. These parameters now have to be tuned in such a way that terms proportional to $\psi \Nabla^4 \psi$ (of order $\varepsilon_\mathrm{gr}^4$) and $\phi\psi$ (of order $\varepsilon_\mathrm{cor}$) are kept in the free energy, whereas terms proportional to $\psi \Nabla^6 \psi$ (of order $\varepsilon_\mathrm{gr}^6$) and $\phi \Nabla^2 \psi$ (of order $\varepsilon_\mathrm{gr}^2\varepsilon_\mathrm{cor}$) are dropped. This can be achieved considering the distinctive limit characterized by the condition 
\begin{equation}
\mathcal{O}(\varepsilon_\mathrm{cor}) = \mathcal{O}(\varepsilon_\mathrm{gr}^4).  
\end{equation}
To elucidate the physical meaning of the smallness parameter $\varepsilon_\mathrm{cor}$, we consider the random phase approximation \cite{EmmerichEtAl2012}
\begin{equation}
c_{ij}(\vec{r}-\vec{r}_1) = - \beta U_{ij}(\vec{r}-\vec{r}_1),
\end{equation}
where $U_{ij}$ is the interaction potential between particles of species $i$ and $j$. Thus, in the simplest case, the direct correlation function depends on the strength of the interaction, i.e., the assumption \eqref{scaling} implies that the interaction of the particles of different species is weaker than between particles of the same species. This would imply that model 1 describes a mixture of active and passive particles where the interaction between different particle types is weaker than the interactions of the particles of one species among themselves.

The reason for why this argument is somewhat naive is that the quantitative predictions obtained for the PFC model parameters by a derivation from DFT can be quite inaccurate, since the assumptions made in this derivation ($\psi$ and $\phi$ are slowly varying in space) are not really justified in the case of crystallization \cite{JaatinenAEA2009,teVrugtNS2022}. Nevertheless, one can still learn something from such derivations, since the qualitative predictions and the mathematical structure of the PFC models so obtained can still be quite accurate \cite{ArcherRRS2019}, and it is the qualitative structure that we are interested in in the present work. However, it should be noted that in practice, one typically obtains the coefficients of a PFC model by fitting a fourth-order polynomial to the Fourier-transformed direct correlation function $\hat{c}(\vec{k})$ \cite{JaatinenAEA2009} depending on the wavenumber $\vec{k}$. Consequently, dropping the gradients in the linear coupling is a good approximation as long as it is a good approximation to replace $\hat{c}_{\psi\phi}(\vec{k})$ by a constant while $\hat{c}_{\psi\psi}(\vec{k})$ and $\hat{c}_{\phi\phi}(\vec{k})$ are fitted with fourth-order polynomials. This, however, is consistent with the basic conclusion of our \ZT{naive} argument, namely that we can drop the gradient terms if $c_{\psi\phi}$ is small compared to the other correlation functions. The reason for this is that, if $c_{\psi\phi}$ is small, any errors we make in fitting this function (such as replacing it by a constant) are less relevant for the overall accuracy of the free energy functional we derive.

The direct correlation functions (more precisely: the static structure factors $S$ which can be calculated from them) are required as an input not only in PFC models, but also in the mode coupling theory of the glass transition \cite{Das2004}. A mode-coupling theory for a mixture of active and passive particles has recently been developed by \citet{FengH2018}, who obtained the structure factors via Brownian dynamics simulations. They found that, while the overall shape of the partial structure factor $S_{\psi\phi}(k)$ ($k =\norm{\vec{k}}$) is similar to that of the partial structure factors $S_{\psi\psi}(k)$ and $S_{\phi\phi}(k)$, $S_{\psi\phi}(k)$ is smaller (around zero and even negative for some $k$). 
 
The Fourier-transformed direct correlation functions are related to the structure factors by \cite{FengH2018}
\begin{equation}
\begin{pmatrix}
\rho_\psi \hat{c}_{\psi\psi} & \sqrt{\rho_\psi\rho_\phi}\hat{c}_{\psi\phi}\\
\sqrt{\rho_\psi\rho_\phi}\hat{c}_{\psi\phi} &\rho_\phi \hat{c}_{\phi\phi}
\end{pmatrix}   
=\Eins - 
\frac{\begin{pmatrix}
S_{\phi\phi} &  -S_{\psi\phi}\\
-S_{\psi\phi}  &S_{\psi\psi}
\end{pmatrix}}{S_{\psi\psi}S_{\phi\phi}-S_{\psi\phi}^2}.
\label{matrixequation}
\end{equation}

Equation \eqref{matrixequation} shows that, if $S_{\psi\phi}$ is small, then (other things being equal) $\hat{c}_{\psi\phi}$ will also be small. Consequently, the simulation results from Ref.\ \cite{FengH2018} support the assumption made in model 1 that gradient terms in the linear coupling are less important than the other gradient terms.
 
Dropping the gradient terms in the coupling in \cref{freeenergymodel2a} thus gives the free energy of \textit{model 1a}, i.e.,
\begin{align}
\nonumber F &= \INT{}{}{^2r}\bigg(\frac{1}{2}\psi(\epsilon_\psi + (q_\psi^2 + \Nabla^2)^2)\psi + \frac{1}{4}\psi^4\\ 
&\quad\,+\frac{1}{2}\phi(\epsilon_\phi + G_\phi(q_\phi^2 + \Nabla^2)^2)\phi + \frac{1}{4}\phi^4 + a_1\psi \phi \label{freeenergymodel1a}\\
\nonumber &\quad\,+ \frac{1}{2}C_1 \vec{P}^2 + \frac{1}{4}C_2\vec{P}^4 + C_3\psi\vec{P}^2 + C_4 \psi^2\vec{P}^2\bigg)
\end{align}
with the linear coupling parameter
\begin{equation}
a_1 = \epsilon_{\mathrm{coup}} + G_{\mathrm{coup}}q_{\mathrm{coup}}^4.
\end{equation}
Model 1a still has nonlinear terms proportional to $\psi\vec{P}^2$, $\psi^2\vec{P}^2$, and $\vec{P}^4$ that are not present in model 1. Compared to the interaction-dependent terms, it is more difficult to see how one can eliminate them given that they arise also from the ideal gas free energy \eqref{idealgasfreeenergy}. Thus, it looks as if they are present regardless of the assumptions we make about interactions. 

However, the situation is more intricate. If we make the assumption that the interaction does not depend on the particle orientation as it is the case for Brownian spheres, in the passive limit, the free energy should not depend on $\vec{P}$. In the active case, we can add a term $\frac{1}{2}C_1\vec{P}^2$ with $C_1 > 0$ to the free energy to account for orientational diffusion (for orientation-independent interactions, we always have $C_1 > 0$ since $C_1$ then only arises from the ideal gas term). This is unproblematic for the passive limit since, for $v_0 = 0$, $\vec{P}$ does not couple to $\psi$. Thus, to ensure the correct passive limit, we should (depending on other choices we make in modeling the interactions) use model 3c, 2c, or 1c.

What is confusing here is that the limit $v_0 \to 0$ based on models of type a does not give us a model of type c even for orientation-independent interactions, since the terms nonlinear in $\vec{P}$ and, in particular, the terms coupling $\psi$ and $\vec{P}$ are still present. In the ideal gas limit, passive models of type a give
\begin{align}
\partial_t \psi &= \Nabla^2(\epsilon_\psi \psi + \psi^3 + C_3 \vec{P}^2 
+ 2 C_4 \psi \vec{P}^2),\label{igl1}\\
\partial_t \vec{P} &= (\Nabla^2 - D_r) (C_1 \vec{P} + C_2 \vec{P}^3 + 2C_3 \psi \vec{P} + 2C_4 \psi^2\vec{P}).
\label{igl}
\end{align}
Note that taking the ideal gas limit is problematic in the rescaled model also because we have forced the prefactor of the term $\psi\Nabla^4\psi/2$ in the free energy, which should be zero in the ideal gas limit, to be one. We have simply dropped this term in \cref{igl1,igl}, but we now will turn to the model given by \cref{psidim,pdim,phidim} (i.e., the \ZT{original} model without rescaling and nondimensionalization), to discuss the ideal gas limit.

Equations \eqref{igl1} and \eqref{igl} show that there is still a coupling between $\vec{P}$ and $\psi$ (which should not be the case for spheres in the passive limit), and also some nonlinear terms which are obviously unphysical. Physically, however, we should simply have a linear diffusion equation for $\psi$. This is also the result that we would get from DDFT.  To see the origin of this problem, it is helpful to take one step back and consider the single-component PFC model for passive spheres. This was discussed in much detail by \citet{ArcherRRS2019}, who identified the Taylor expansion of the logarithm and the constant mobility approximation (CMA) as crucial steps that lead to potential unphysical behavior. Here, we will briefly discuss the implications of this issue for the dynamics. For the passive field $\phi$, the model \eqref{phidim} with CMA and free energy \eqref{idealgasfreeenergy} gives
\begin{equation}
\partial_t\phi = D\Nabla^2 \bigg(\phi - \frac{\phi^2}{2} + \frac{\phi^3}{3}\bigg).  
\label{phiidealgas}
\end{equation}
The result \eqref{phiidealgas} is not correct since, for an ideal gas, the dynamics of $\phi$ is simply given by the standard diffusion equation. If we use the nonconstant mobility $1+\psi$ obtained from DDFT instead, we find
\begin{align}
\nonumber \partial_t \phi &= D\Nabla\cdot \bigg((1+\phi)\Nabla\bigg(\phi -\frac{\phi^2}{2}+ \frac{\phi^3}{3}\bigg)\!\bigg)\\
&= D\Nabla^2\bigg(\phi + \frac{\phi^2}{2} - \frac{\phi^2}{2}- \frac{\phi^3}{3} + \frac{\phi^3}{3} \bigg)\\
\nonumber &= D\Nabla^2 \phi,
\end{align}
having used $\Nabla\cdot(\phi\Nabla\phi) = \Nabla^2\phi^2/2$, $\Nabla\cdot(\phi \Nabla \phi^2)/2 = \Nabla^2\phi^3/3$ and dropping terms of order $\phi^4$. This is the correct description of the dynamics. If, however, we make a CMA, the only way to recover the diffusion equation for noninteracting particles is to set $F=\beta^{-1}\bar{\rho}_\phi\TINT{}{}{^2r}\phi^2/2$.

Similarly, if we describe the field $\psi$ with CMA using \cref{psidim}, insert the ideal gas free energy \eqref{idealgasfreeenergy}, and set $v=0$, we find
\begin{equation}
\partial_t \psi = D_0\Nabla^2\bigg(\psi - \frac{\psi^2}{2} + \frac{\psi^3}{3} - \frac{\vec{P}^2}{4} + \frac{\psi\vec{P}^2}{2}\bigg).
\label{psiidealgas}
\end{equation}
Without a CMA one finds\footnote{To get the first equality of \cref{psiwithoutcma}, replace $D_0\Nabla^2$ by $D_0\Nabla\cdot(1+\psi + \vec{P}\cdot\uu)\Nabla$ in \cref{firsteqapprox} (with $v=0$) and integrate over $\varphi$. The term $(\Nabla\delta F/\delta\vec{P})\cdot\vec{P}$ is, in index notation with summation convention, given by $(\partial_i\delta F/\delta P_j)P_j$.}
\begin{align}
\nonumber \partial_t\psi &= \frac{D_0\beta}{2\pi\bar{\rho}_\psi}\Nabla\cdot\bigg((1+\psi)\Nabla\Fdif{F}{\psi} + \bigg(\Nabla\Fdif{F}{\vec{P}}\bigg)\cdot\vec{P}\bigg)\\
\nonumber &=D_0\Nabla\cdot\bigg((1+\psi)\Nabla\bigg(\psi - \frac{\psi^2}{2} + \frac{\psi^3}{3} - \frac{\vec{P}^2}{4} + \frac{\psi\vec{P}^2}{2}\bigg)\\
&\quad\, + \vec{P}\cdot\Nabla\bigg(\frac{\vec{P}}{2} - \frac{\psi\vec{P}}{2} + \frac{\psi^2\vec{P}}{2}+\frac{\vec{P}^3}{8}\bigg)\!\bigg)\label{psiwithoutcma}\\
\nonumber &=D_0\Nabla^2\psi,
\end{align}
where in the last step we have ignored terms of fourth order in the fields. Here, terms coupling $\psi$ and $\vec{P}$ disappear as they should. 

One might now ask why we do not just stop the Taylor expansion of the logarithm after the quadratic term, in which case the PFC model gives the correct equation of motion for the ideal gas limit (i.e., the diffusion equation). The reason is that we also have to ensure the correct static results in the passive limit: For the passive PFC model, as for the passive DDFT, the system evolves toward the state that minimizes the free energy functional. Consequently, whether the phase transitions predicted by the model are correct depends on the accuracy of the free energy functional, which increases if we take into account higher orders in the expansion of the logarithm. 

A particularly important reason why the higher-order terms can be important is \textit{stabilization}. As an example, let us assume that we can treat interactions in the Ramakrishnan-Yussouff approximation. In this case, the coefficient $C_1$ has contributions from the ideal gas and excess free energy, while the coefficient $C_2$ arises solely from the ideal gas free energy. Consequently, although we certainly have $C_2 > 0$, we might have $C_1 < 0$ if the interactions are sufficiently strong. In this case, if we drop higher-order terms in the free energy, the polarization would grow without bounds, which is obviously unphysical. This can be prevented by introducing a term $\frac{1}{4}C_2\vec{P}^4$, resulting from the ideal gas free energy. In this case, instead of an (unphysical) blow-up, we get the well known phenomenon of self-polarization \cite{OphausGT2018}, which, as is evident from the microscopic derivation, is only possible for orientation-dependent interactions. Consequently, if the terms in \cref{freeenergymodel3a} that are quartic in the free energy become negative due to interactions, it might be necessary to include even more terms in the ideal gas expansion.

For this particular purpose, however, it is sufficient to keep the term $\frac{1}{4}C_2\vec{P}^4$ in addition to $\frac{1}{2}C_1 \vec{P}^2$. Terms coupling $\psi$ and $\vec{P}$, which (in the Ramakrishnan-Yussouff approximation) only arise from the ideal gas free energy, are almost always ignored in PFC models. While this is typically done without any justification, we can now give a physical motivation for such models: They are a compromise between models of type a, where all nonlinear terms in $\vec{P}$ are kept (giving a stable and accurate free energy functional and an incorrect diffusion equation), and models of type c, which are appropriate for active Brownian spheres. Since they are in between these two extremes, we call them \ZT{models of type b}. Type b models cannot be obtained via a systematic expansion in a smallness parameter, they rather arise as a minimal model allowing for orientation-dependent interactions. They also come in three forms, with the most complicated one being \textit{model 3b}:
\begin{align}
\nonumber F &= \INT{}{}{^2r}\bigg(\frac{1}{2}\psi(\epsilon_\psi + (q_\psi^2 + \Nabla^2)^2)\psi + \frac{1}{4}\psi^4\\ 
\nonumber &\quad\,+\frac{1}{2}\phi(\epsilon_\phi + G_\phi(q_\phi^2 + \Nabla^2)^2)\phi + \frac{1}{4}\phi^4 \\
&\quad\,+ \psi(\epsilon_{\mathrm{coup}} + G_{\mathrm{coup}}(q_{\mathrm{coup}}^2 + \Nabla^2)^2)\phi \label{freeenergymodel3b}\\
\nonumber &\quad\,+ \frac{1}{2}C_1 \vec{P}^2 + \frac{1}{4}C_2\vec{P}^4 + a_2 \psi^2\phi + a_3 \psi \phi^2\\
\nonumber &\quad\,+ a_4 \psi^3\phi + a_5 \psi^2\phi^2 + a_6 \psi\phi^3\bigg).
\end{align}
Terms coupling $\phi$ and $\vec{P}$ have also been dropped (since it is somewhat inconsistent to keep them if we drop terms coupling $\psi$ and $\vec{P}$). A simpler variant, obtained using a Ramakrishnan-Yussouff approximation, is \textit{model 2b}:
\begin{align}
\nonumber F &= \INT{}{}{^2r}\bigg(\frac{1}{2}\psi(\epsilon_\psi + (q_\psi^2 + \Nabla^2)^2)\psi + \frac{1}{4}\psi^4\\ 
\nonumber &\quad\,+\frac{1}{2}\phi(\epsilon_\phi + G_\phi(q_\phi^2 + \Nabla^2)^2)\phi + \frac{1}{4}\phi^4 \label{freeenergymodel2b}\\
&\quad\,+\psi(\epsilon_{\mathrm{coup}} + G_{\mathrm{coup}}(q_{\mathrm{coup}}^2 + \Nabla^2)^2)\phi\\
\nonumber &\quad\,+ \frac{1}{2}C_1 \vec{P}^2 + \frac{1}{4}C_2\vec{P}^4\bigg).
\end{align}
Finally, by dropping gradients, we can also get \textit{model 1b}:
\begin{align}
\nonumber F &= \INT{}{}{^2r}\bigg(\frac{1}{2}\psi(\epsilon_\psi + (q_\psi^2 + \Nabla^2)^2)\psi + \frac{1}{4}\psi^4\\ 
&\quad\,+\frac{1}{2}\phi(\epsilon_\phi + G_\phi(q_\phi^2 + \Nabla^2)^2)\phi + \frac{1}{4}\phi^4 \label{freeenergymodel1b}\\
\nonumber &\quad\,+ a_1\psi \phi+ \frac{1}{2}C_1 \vec{P}^2 + \frac{1}{4}C_2\vec{P}^4\bigg).
\end{align}
Let us now assume that there are no orientation-dependent interactions at all. In this case, $C_1$ is guaranteed to be positive, and there is no need for terms of higher order in $\vec{P}$ to ensure stabilization. On the other hand, as discussed above, these terms give unphysical results in the ideal gas limit. If we, based on these considerations, drop all terms involving $\vec{P}$ apart from $C_2\vec{P}^2$ in the free energy \eqref{freeenergymodel3a} of model 3a as it is appropriate for active Brownian spheres, we are left with \textit{model 3c}:
\begin{align}
\nonumber F &= \INT{}{}{^2r}\bigg(\frac{1}{2}\psi(\epsilon_\psi + (q_\psi^2 + \Nabla^2)^2)\psi + \frac{1}{4}\psi^4\\ 
\nonumber &\quad\,+\frac{1}{2}\phi(\epsilon_\phi + G_\phi(q_\phi^2 + \Nabla^2)^2)\phi + \frac{1}{4}\phi^4 \\
&\quad\,+ \psi(\epsilon_{\mathrm{coup}} + G_{\mathrm{coup}}(q_{\mathrm{coup}}^2 + \Nabla^2)^2)\phi\label{freeenergymodel3c}\\
\nonumber &\quad\,+ \frac{1}{2}C_1 \vec{P}^2 + a_2 \psi^2\phi + a_3 \psi \phi^2\\
\nonumber &\quad\,+ a_4 \psi^3\phi + a_5 \psi^2\phi^2 + a_6 \psi\phi^3\bigg).
\end{align}
Also making a Ramakrishnan-Yussouff approximation then gives \textit{model 2c}:
\begin{align}
\nonumber F &= \INT{}{}{^2r}\bigg(\frac{1}{2}\psi(\epsilon_\psi + (q_\psi^2 + \Nabla^2)^2)\psi + \frac{1}{4}\psi^4\\ 
\nonumber &\quad\,+\frac{1}{2}\phi(\epsilon_\phi + G_\phi(q_\phi^2 + \Nabla^2)^2)\phi + \frac{1}{4}\phi^4 \\
&\quad\,+ \psi(\epsilon_{\mathrm{coup}} + G_{\mathrm{coup}}(q_{\mathrm{coup}}^2 + \Nabla^2)^2)\phi\label{freeenergymodel2c}\\
\nonumber &\quad\,+ \frac{1}{2}C_1 \vec{P}^2\bigg).
\end{align}
Finally, if we also drop gradient terms in the interaction, we get the simple \textit{model 1c}:
\begin{align}
\nonumber F &= \INT{}{}{^2r}\bigg(\frac{1}{2}\psi(\epsilon_\psi + (q_\psi^2 + \Nabla^2)^2)\psi + \frac{1}{4}\psi^4\\ 
&\quad\,+\frac{1}{2}\phi(\epsilon_\phi + G_\phi(q_\phi^2 + \Nabla^2)^2)\phi + \frac{1}{4}\phi^4 \label{freeenergymodel1c}\\
\nonumber &\quad\,+ a_1\psi \phi+ \frac{1}{2}C_1 \vec{P}^2\bigg).
\end{align}

Model 1 introduced in \cref{equations} is identical to model 1c apart from the fact that $G_\phi \neq 1$ and $M_\phi \neq 1$ in model 1b. In principle, model 1 is \ZT{only} a special case of model 1c recovered by setting the coefficients $G_\phi$ and $M_\phi$ to one. Recall that we can get $M_\phi=1$ also by adjusting the reference densities (see \cref{mphi2}). Setting $M_\phi=1$ and $G_\phi = 1$ and dropping the subscript \ZT{1} from $a_1$, \cref{psi4,p4,phi4,freeenergymodel1c} give
\begin{align}
\partial_t \psi &= \Nabla^2\frac{\delta F}{\delta \psi} - v_0 \Nabla \cdot \vec{P},\\ 
\partial_t \vec{P}&= \Nabla^2\frac{\delta F}{\delta \vec{P}} - D_r \frac{\delta F}{\delta \vec{P}} - v_0 \Nabla\psi,\\
\partial_t \phi &= \Nabla^2\frac{\delta F}{\delta \phi},\\
\nonumber F &= \INT{}{}{^2r} \Big( \frac{1}{2}\psi(\epsilon_\psi + (q_\psi^2 + \Nabla^2)^2)\psi + \frac{1}{4}\psi^4\\ 
&\quad\,+\frac{1}{2}\phi(\epsilon_\phi + (q_\phi^2 + \Nabla^2)^2)\phi + \frac{1}{4}\phi^4 \label{freeenergymodel1}\\
\nonumber &\quad\,+ a\psi \phi+ \frac{1}{2}C_1 \vec{P}^2+\frac{1}{4}C_2 \vec{P}^4 \Big).
\end{align}
To summarize: Model 1 is a PFC model for a mixture of active and passive particles, which assumes that
\begin{itemize}
\item nonlinearities in $\vec{P}$ resulting from the ideal gas free energy can be ignored (generalization: models 3a, 3b, 2a, 2b, 1a, and 1b);
\item the excess free energy functional can be obtained from the Ramakrishnan-Yussouff approximation (generalization: models 3a-3c);
\item gradient terms in the coupling can be ignored (generalization: models 3a-3c and 2a-2c);
\item the mobilities of both fields are equal;
\item the factor $G_\phi$ can be set to one (i.e., the prefactor of the gradient terms is equal for both fields).
\end{itemize}
We have thus achieved a systematic derivation of the minimal model 1 and identified the approximations required for obtaining it. Moreover, we have derived and classified more general models (3a-3c, 2a-2c, and 1a-1c), which provide a very general description of mixtures of active and passive particles. An overview over the model hierarchy is given in \cref{overview}.

\section{\label{linear}Linear stability analysis}
Next, we consider the linear stability of uniform states for a mixture of active and passive particles using the above derived model~1 as given by \cref{uno,tres,dos}. The limiting case of a mixture of passive particles can easily be obtained by setting $v_0 = 0$. From now on we restrict ourselves to one spatial dimension.
 
Introducing the notation $\StatesVector=(\psi, P, \phi)^{\mathrm{T}}$ for the tupel of fields,
uniform steady states $\StatesVector^\star=(\psi^*, P^*, \phi^*)^{\mathrm{T}}$ are characterized by $\partial_t\StatesVector=\FieldTupel{0}$. As in \cref{uno,tres,dos} the mean densities are explicitly included as parameters $\bar{\psi}$ and $\bar{\phi}$, we have $\psi^*=0$ and $\phi^*=0$. Furthermore, $P^*=0$ is the only homogeneous stationary solution of Eq.~\eqref{tres}. To obtain the linear stability of the uniform states, \cref{uno,tres,dos} are linearized in small perturbations about $(0, 0, 0)^{\mathrm{T}}$ using the ansatz 
\begin{equation}
\StateVector
	= \FieldTupel{\delta\StateVector}\,
	e^{\ii kx + \lambda t}
\end{equation}
with growth rate $\lambda$. This yields
\begin{equation}
	\lambda\,\FieldTupel{\delta\StateVector}
	=  \underline{J}\, \FieldTupel{\delta\StateVector}
\end{equation}%
with the matrix 
\begin{equation}
\underline{J} = \begin{pmatrix}
	    J_\psi & - \ii v_0 k & -ak^2\\
	    - \ii v_0 k & J_\text{p} & 0 \\
	    -ak^2 & 0 & J_\phi
	\end{pmatrix},
\end{equation}
where
\begin{equation}
J_{\chi}(k)  \equiv - k^2\big(\epsilon_{\chi} + (q_{\chi}^2 - k^2)^2 + 3(\chi^*)^2\big)
\end{equation} 
for $\chi=\psi,\phi$ and
\begin{equation}
J_\text{p}(k)  \equiv - C_1k^2 - D_\text{r}C_1.
\end{equation} 

The condition $\det(\underline{J} - \lambda\underline{\mathds{1}}) = 0$ leads to the cubic equation
\begin{equation}
-\lambda^3 + \alpha_1\lambda^2 + \alpha_2\lambda + \alpha_3 = 0,
\label{cubic}
\end{equation}
where
\begin{align}
\alpha_1(k) & \equiv J_\psi + J_\phi + J_\text{p}, \\
\alpha_2(k) & \equiv - \big((J_\psi + J_\phi)J_\text{p} + J_\psi J_\phi + v_0^2k^2 - a^2k^4\big), \\
\alpha_3(k) & \equiv J_\psi J_\phi J_\text{p} + v_0^2k^2J_\phi - a^2k^4J_\text{p}.
\end{align}
The solution of the cubic equation \eqref{cubic} can be determined by Cardano's method. Due to the complexity of the analytical solution, the equation is numerically solved using Python. 

\begin{figure}
\centering
\includegraphics[width=\linewidth]{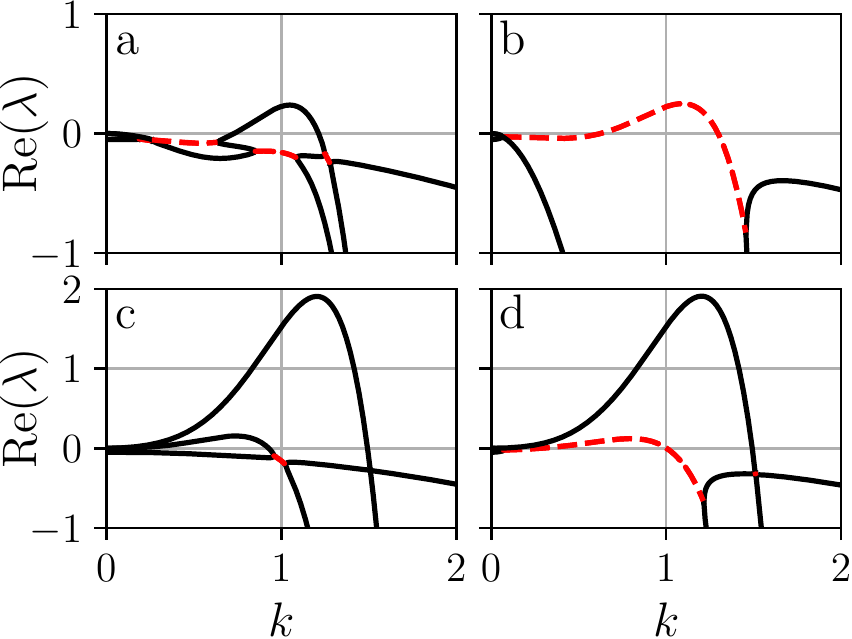}
\caption{\label{fig:disprel}Dispersion relations with black solid [red dashed] lines corresponding to real [complex] eigenvalues. The dispersion relations presented in (a) and (b) show instabilities similar to the one-component active PFC model \cite{OphausGT2018}: (a) gives a monotonic small-scale instability at $\bar\psi = \bar\phi = -0.7$, $ q_\psi=q_\phi = 1 $, and $v_0 = 0.05$; (b) presents an oscillatory small-scale instability at $\bar\psi = -0.55$, $\bar\phi = -1.5$, $ q_\psi=q_\phi = 1 $, and $v_0 = 0.4$; (c) exhibits two monotonic instability modes at $\bar\psi = -0.6$, $\bar\phi = 0$, $ q_\psi=0.5$, $q_\phi = 1 $, and $v_0 = 0.05$; (d) shows a case with a monotonic and an oscillatory instability mode at $\bar\psi = -0.5$, $\bar\phi = 0$, $ q_\psi=0.5$, $q_\phi = 1 $, and $v_0 = 0.4$. The remaining parameters are $ \epsilon_\psi=\epsilon_\phi=-1.5 $, $ a = -0.2 $, $ D_r = 0.5$, and $ C_1 = 0.1 $.}
\end{figure}
 
The real parts $\textrm{Re}(\lambda)$ of dispersion relations $\lambda_j(k)$ with $j=1,2,3$ are displayed in \cref{fig:disprel} for four exemplary parameter choices. In \cref{fig:disprel}(a), a case of low activity $v_0=0.05$ is given where a monotonic small-scale instability occurs, i.e., above instability onset, where a single mode of wavenumber $k_\mathrm{c}\neq0$ (the critical value) becomes unstable, a finite band of unstable wavenumbers exists centered about $k_\mathrm{c}$.  In a direct time simulation this produces a resting crystal (not shown). A decrease in the effective temperature $\epsilon_\psi$ will widen the band of unstable wavenumbers. A case of higher activity $v_0=0.4$ is presented in \cref{fig:disprel}(b). Again, a small-scale instability arises (finite band of unstable $k$ centered about  $k_\mathrm{c}\neq0$), however, the instability is now oscillatory. A time simulation produces a traveling periodic state, i.e., a traveling crystal (not shown). Both cases presented in Figs.\ \ref{fig:disprel}(a) and \ref{fig:disprel}(b) are also encountered in one-component active PFC models \cite{OphausGT2018}. This does not apply to the remaining cases of \cref{fig:disprel}. In particular, \cref{fig:disprel}(c) shows a case of small activity $v_0=0.05$ where two monotonic small-scale modes are unstable. This is also found in the limiting case of a PFC model for a mixture of passive particles \cite{HollAT2020}. Increasing the activity to $v_0=0.4$ renders one of these unstable modes oscillatory, as presented in \cref{fig:disprel}(d). 

Note that an increase in the linear coupling between the densities $\psi$ and $\phi$ generally results in a further destabilization. This can already be anticipated inspecting the dispersion relation in the passive limit that reads \cite{HollAT2020} 
\begin{align}
\lambda_\pm (k) = \frac{J_\psi + J_\phi \pm \sqrt{(J_\psi - J_\phi)^2 + 4a^2k^4}}{2}.
\end{align}
There, increasing the coupling strength $a$, always results in an increase of the largest eigenvalue $\lambda_+$.

\section{\label{nonlinear}Nonlinear states} 
In this section, we present selected fully nonlinear states as obtained by numerical path continuations \cite{KrauskopfOG2007,EngelnkemperGUWT2019}, as well as results of direct time simulations. Thereby, the focus is on effects that are not seen in the standard one-component active PFC model. For the path continuations we employ the package pde2path \cite{UeckerWR2014} while all time simulations are performed using a semi-implicit Euler scheme with adaptive step size. All results from time simulations are presented after initial transients have decayed.
In the following, we only consider the particular case of equal temperature parameters $\epsilon_\psi = \epsilon_\phi$, and equal critical wavenumbers $q_\psi = q_\phi = 1$ for the active and passive particles, and drop the subscripts accordingly.
	
\begin{figure}
\centering
\includegraphics[width=\linewidth]{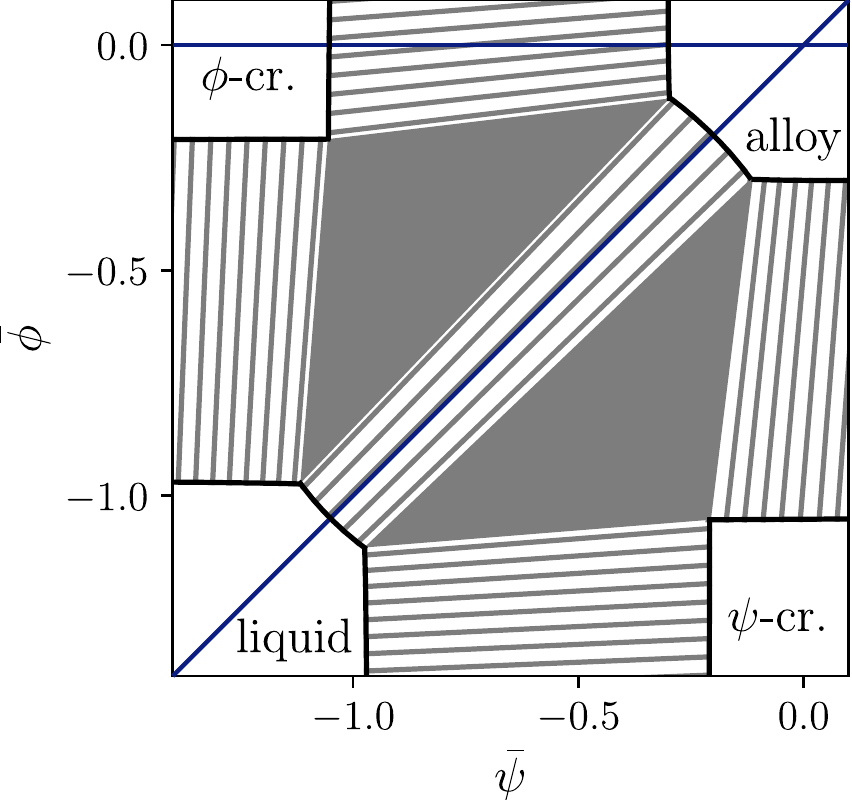}
\caption{\label{fig:phasediagram}Phase diagram of the PFC model for a mixture of passive particles in 1D displayed in the $(\bar{\psi},\bar{\phi})$-plane at $\epsilon = -1.5$ and $v_0 = 0$. 
Four phases (liquid, alloy, $\psi$-crystal, $\phi$-crystal) can be distinguished. 
The binodal lines are marked in black. Coexisting states on the binodals are connected by gray tie lines. The thus hatched areas correspond to two-phase coexistence of adjacent phases. Three-phase coexistence is indicated by triangular gray shaded areas. The binodals and tie lines are only presented for thermodynamic coexistence, not for metastable or linearly unstable coexistence. Along the horizontal blue line, the bifurcation diagrams in \cref{fig:bd_v0023_horz,fig:bd_v003_horz,fig:bd_v004_horz} are presented, while the diagonal blue line denotes the path corresponding to \cref{fig:bd_v003_diag}. The remaining parameters are as in \cref{fig:disprel}.}
\end{figure}
	
\begin{figure}
\includegraphics[width=\linewidth]{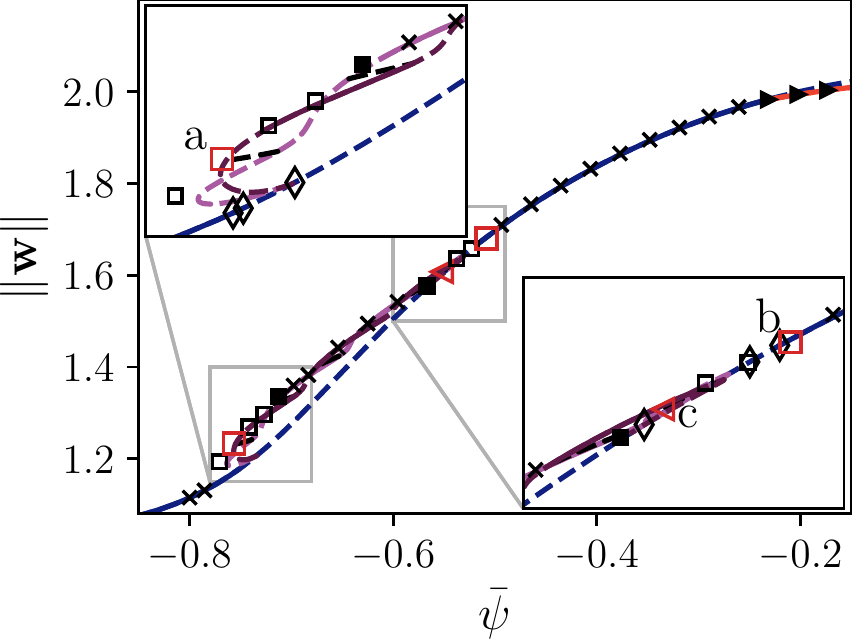}
\caption{\label{fig:bd_v0023_horz}Bifurcation diagram showing branches of one-dimensional states of the PFC model 1 \eqref{uno}-\eqref{dos} for coupled active and passive particles. Shown is the L$^2$ norm $\norm{\StatesVector}$ as a function of the mean density $ \bar\psi $ at fixed $ \bar\phi = 0 $ and activity $v_0=0.23$. Solid [dashed] lines indicate stable [unstable] states. The blue line corresponds to crystalline (periodic) states with $n=8$ peaks. The intertwined light and dark purple lines represent the slanted snaking of branches of LSs with odd and even number of peaks, respectively. They are interconnected by branches of asymmetric states (black lines). The branch of steadily traveling periodic states is given as orange line. 
The various states obtained by time simulations are indicated by symbols according to Table \ref{tab:statemarkers}.
The insets magnify regions where the branch of resting periodic states changes stability. Hopf bifurcations occurring on this branch are marked by diamonds. Figure \ref{fig:sols_v0023_horz} shows selected space-time plots at the locations marked by letters ``a''-``c''. The corresponding symbols are enlarged and red. 
The remaining parameters are $ \epsilon=-1.5 $, $ q = 1 $, $ a = -0.2 $, $ D_r = 0.5$, and $ C_1 = 0.1 $. The domain size is fixed at $ L= 16 \pi$.}
\end{figure}

\begin{figure}
\centering
\includegraphics[width=\linewidth]{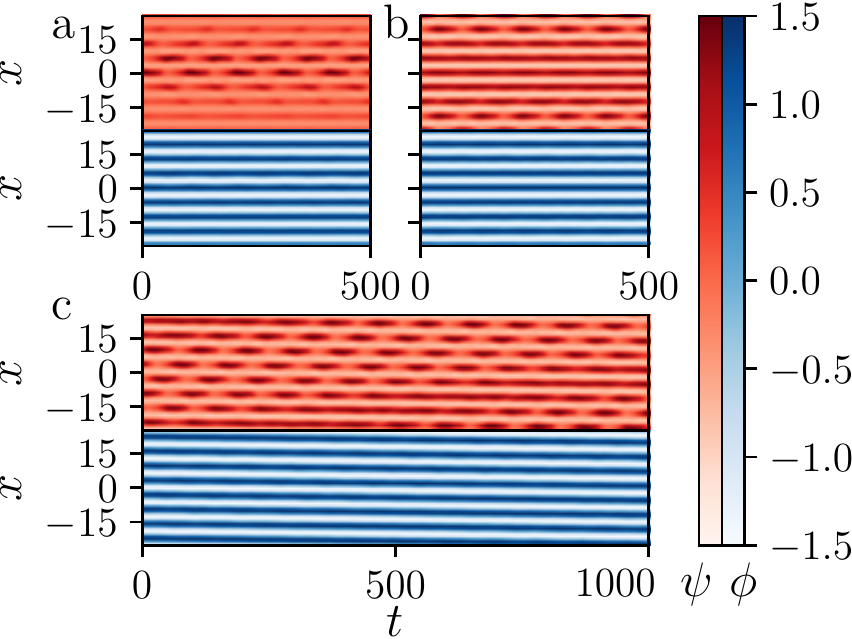}
\caption{\label{fig:sols_v0023_horz}Panels (a) to (c) show space-time plots of the densities of active ($\psi(x,t)$) and passive ($\phi(x, t)$) particles at the locations indicated by corresponding letters in \cref{fig:bd_v0023_horz} (after transients have decayed). Panels (a) and (b) show alternating LSs in the interval $ t = [0, 500] $. Due to the slower dynamics, the interval in panel (c) is $ t = [0, 1000] $, showing alternating traveling LSs.}
\end{figure}

\begin{figure}
\includegraphics[width=\linewidth]{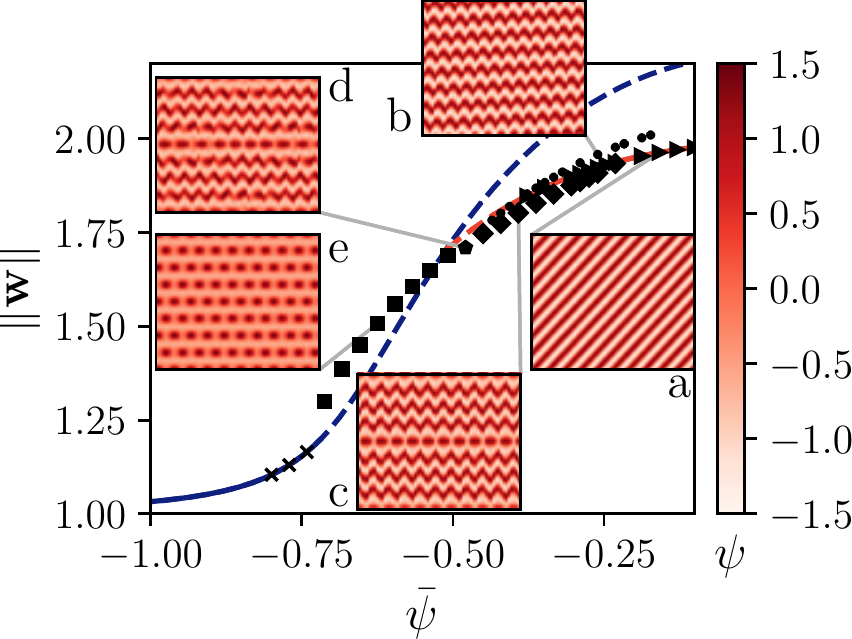}
\caption{\label{fig:bd_v003_horz}Bifurcation diagram as a function of the mean density $ \bar\psi $ at fixed activity $v_0=0.3$. 
The  model, solution measure, remaining parameters, symbols, and line styles are as in Fig.~\ref{fig:bd_v0023_horz}, while the insets show space-time plots as in \cref{fig:sols_v0023_horz}(a), focusing on the density of the active particles $\psi$.}
\end{figure}

\begin{figure}
\includegraphics[width=\linewidth]{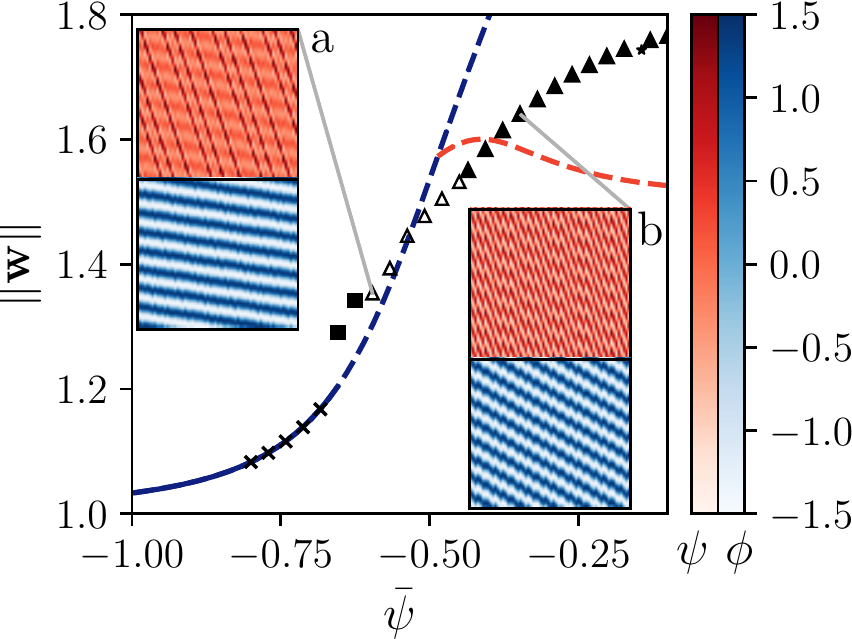}
\caption{\label{fig:bd_v004_horz}Bifurcation diagram as a function of the mean density $ \bar\psi $ at fixed activity $v_0=0.4$. The two insets show space-time plots of the active and passive densities (upper panel: $\psi(x,t)$; lower panel: $\phi(x,t)$) at the marked locations in an interval $ t = [0, 500]$ after initial transients have decayed. The model, solution measure, remaining parameters, symbols, and line styles are as in Fig.~\ref{fig:bd_v0023_horz}, while the insets are as in Fig.\ \ref{fig:sols_v0023_horz}(a).}
\end{figure}

\begin{figure}
\includegraphics[width=\linewidth]{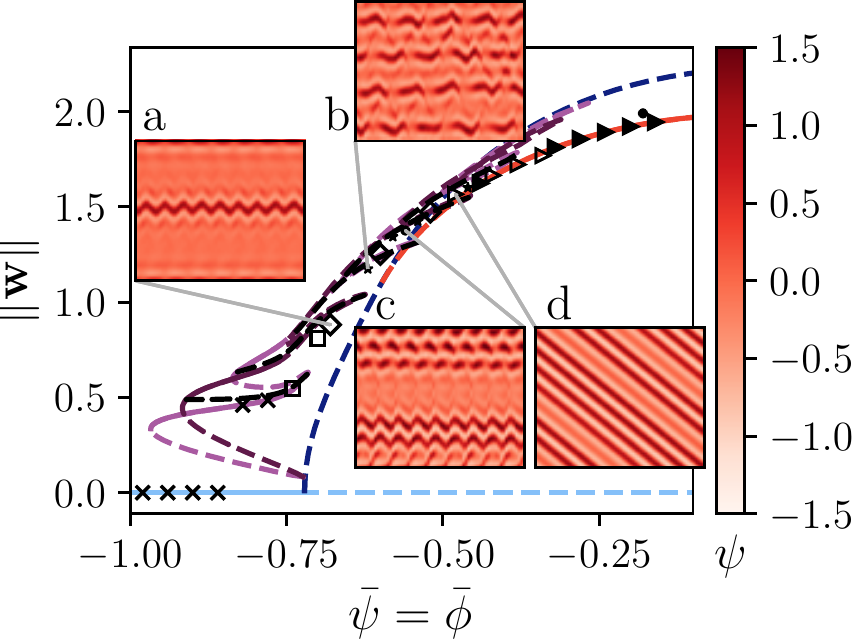}
\caption{\label{fig:bd_v003_diag}Typical bifurcation diagram as a function of the mean densities $ \bar\psi = \bar{\phi}$ at fixed activity $v_0=0.3$.
The light blue line corresponds to liquid (uniform) states. The  model, solution measure, remaining parameters, symbols, and line styles are as in Fig.~\ref{fig:bd_v0023_horz}, while the insets are as in \cref{fig:sols_v0023_horz}(a), focusing on the density of the active particles $\psi$.}
\end{figure}

First, we determine resulting states in the passive limit $v_0 = 0$. The corresponding phase diagram in \cref{fig:phasediagram} is quite similar to the one presented in Ref.\ \cite{HollAT2020}. Pairs of thermodynamically stable coexisting phases lie on the black binodal lines. Note that, for clarity, in \cref{fig:phasediagram} we only present thermodynamically stable coexistence and omit metastable and linearly unstable coexistence. Particular coexisting states with equal chemical potentials and pressure on two binodals are connected by gray tie lines. Within the hatched region, uniform states are unstable w.r.t.\ phase separation resulting in coexistence of the two bordering phases. The two gray triangles mark regions where phase separation into three phases occurs. Overall, \cref{fig:phasediagram} shows four thermodynamically stable liquid (uniform) and crystal (periodic) phases. At low densities $ \bar\psi $ and $ \bar\phi $, the liquid state is the globally stable state. In the top left and bottom right corners of the shown range, one of $ \bar{\psi} $ and $ \bar{\phi} $ is high while the other one is low, resulting in a crystal state of the particles with the high density, while the other particles are in a weakly modulated liquid state. When $ \bar\psi $ and $ \bar\phi$ are both large, the resulting state is a crystalline alloy: both fields show peaks of similar amplitudes that are in phase. For a deeper discussion of such phase diagrams, see Ref.\ \cite{HollAT2020}.
Using the phase diagram in \cref{fig:phasediagram} for the passive mixture as reference, we analyze the bifurcation behavior in the active case: Fixing the activity at $ v_0 = 0.23, 0.3$, or $0.4$, we
consider two paths through parameter space. First, we vary $\bar\psi$ at fixed $ \bar{\phi} = 0 $  (Figs.~\ref{fig:bd_v0023_horz}-\ref{fig:bd_v004_horz}) and second we consider the path defined by $ \bar{\phi} = \bar{\psi}$ (\cref{fig:bd_v003_diag}).
	
All resting states $\StatesVector = (\psi, P, \phi)^{\mathrm{T}}$ are characterized by their L$^2$ norm 
\begin{equation}
\norm{\StatesVector} = \sqrt{\frac{1}{L}\int_{-L/2}^{L/2} \mathrm{d}x\, (\psi^2 + P^2 + \phi^2)}
\end{equation}
with the domain length $L$. Traveling states are characterized by the temporal average of $\norm{\StatesVector}$.
	
At small activity $ v_0 $, the overall structure of the bifurcation diagrams is very similar to the one in the passive limit (see, e.g., Figs.~8 and 9 of Ref.\ \cite{HollAGKOT2021}). With increasing activity, the parameter range where localized states (LSs) exist slightly decreases \cite{HollAGKOT2021}.

\begin{table}
\caption{\label{tab:statemarkers}The various observed states and the symbols by which they are marked in the bifurcation diagrams in \cref{fig:bd_v0023_horz,fig:bd_v003_horz,fig:bd_v004_horz,fig:bd_v003_diag}.}
\begin{tabular}{|lc|}
\hline
\textbf{State} & \textbf{Marker}\\
\hline
resting (periodic or LS) & $ \times $\\
traveling periodic & $ \blacktriangleright $\\
traveling LS & $ \rhd $\\
traveling periodic, 2 speeds & \UParrow\\
traveling LS, 2 speeds & $ \bigtriangleup $\\
alternating periodic& \ding{110} \\
alternating LS & $ \square $\\
alternating traveling LS& $ \lhd $\\
wiggling, alternating periodic & \ding{117}\\
wiggling, alternating LS & $ \Diamond $\\
wiggling, alternating, 2 periods & \resizebox{0.25cm}{!}{\begin{tikzpicture}
\node[regular polygon,
draw,
regular polygon sides = 5,
fill = black] (p) at (0,0) {};
\end{tikzpicture}}\\
wiggling, traveling periodic & $ \bullet $\\
traveling wiggling LS + alternating LS & $\circ$\\
irregular & $ \star $
\\\hline
\end{tabular}
\end{table}

Here, we first consider states at fixed mean density $ \bar\phi = 0 $ where the field $ \phi $ is in the periodic crystalline state. That is, it provides a periodic background on which, for increasing mean density $ \bar\psi $, LSs, i.e., crystallites in $\psi$, grow. Although the periodic background influences where the active particles prefer to be, it is not entirely static as it is also influenced by the active particles. 
The particular results for $ v_0 = 0.23 $ are summarized in Fig.~\ref{fig:bd_v0023_horz}. At low densities $ \bar\psi $, resting periodic states exist. These are destabilized at $ \bar\psi\approx -0.753 $ in a Hopf bifurcation, marked by the leftmost diamond symbol in the upper left inset of \cref{fig:bd_v0023_horz}. After further destabilization by another Hopf bifurcation at  $ \bar\psi\approx -0.750 $, a subcritical pitchfork bifurcation occurs at $ \bar\psi\approx -0.735$. There, two branches of symmetric LSs emerge with respective odd and even number of peaks.  Both branches of symmetric LSs are initially unstable. On each branch a series of Hopf bifurcations occurs before the first saddle-node bifurcation (not shown) rendering  the LS less unstable. The branches of symmetric LSs are interconnected by ladder branches of asymmetric LSs that emerge in pitchfork bifurcations. As one moves along the branches of symmetric LSs consecutively, pairs of additional peaks are added. Note that, in contrast to the behavior in the passive limit at the same temperature (not shown), this does not involve saddle-node bifurcations. Similar slanted snaking behavior is observed at higher effective temperature in the passive limit of the one-component PFC model (see, e.g., Refs.\  \cite{ThieleARGK2013,HollAGKOT2021}). Between $ \bar\psi = -0.743$ and $\bar{\psi}=-0.537$, both branches of symmetric LSs undergo Hopf and pitchfork bifurcations. As a result, they change stability in such a way that always at least one of them is linearly stable. 

When the whole domain has filled with density peaks, the branches of symmetric LSs reconnect at $ \bar\psi = -0.564 $ to the branch of periodic states. This branch regains stability at $ \bar\psi= - 0.512 $ after undergoing three further Hopf bifurcations, visible in the lower right inset of \cref{fig:bd_v0023_horz}.  Further following the branch, at $ \bar\psi= -0.234 $ it becomes unstable again in a drift pitchfork bifurcation where a branch of steadily traveling periodic states emerges. 
The panels in \cref{fig:sols_v0023_horz} show space-time plots of the density of the active species $ \psi $  and the passive species $\phi$ after transients have decayed. In particular, we show three states that have not yet been described and may not exist for the one-component active PFC model \cite{OphausGT2018,OphausKGT2020}. At $ \bar{\phi} = 0 $, the passive background $ \phi(x,t)$ is periodic, in the considered domain this corresponds to a periodic solution with $ n=8 $ peaks. Figures \ref{fig:sols_v0023_horz}(a) and \ref{fig:sols_v0023_horz}(b) show time-periodic states where the passive background is nearly steady: it only shows small oscillations as reaction to the large amplitude oscillations in the LS in $ \psi $. 
	
Figure \ref{fig:sols_v0023_horz}(a) shows an oscillating LS in $\psi$ on a periodic background in $\phi$. Thereby, each individual peak oscillates like a standing wave, with neighboring peaks oscillating in anti-phase. With other words, active particles alternately co-occupy neighboring sites where passive particles are located. The behavior is strongest at the center of the LS while oscillation amplitudes become smaller as one moves toward the outside tails of the LS. Overall, the structure resembles a spatially localized space-time checkerboard pattern, somewhat similar to the modulated standing waves in Ref.\ \cite{GolovinMBN1999}. We call it an alternating LS. 
 
Figure \ref{fig:sols_v0023_horz}(b) shows an alternating LS at higher $ \bar{\psi} $ that accordingly has a larger spatial extension than the one in Fig.\ \ref{fig:sols_v0023_horz}(a) and nearly fills the domain. Another consequence is the much smaller amplitude of the oscillations at the center of the LS as compared to the larger amplitude at the fringes. With other words, the three central sites of the passive background are always co-occupied by active particles, forming a nearly steady core, while the sites further from the center show oscillating occupancy by active particles. Note that in \cref{fig:bd_v0023_horz} one also finds fully spatially periodic (not localized) variants of such an alternating state. They are marked by filled squares, and a typical space-time plot is given in \cref{fig:bd_v003_horz}(e) (see below).

Figure \ref{fig:sols_v0023_horz}(c) presents a qualitatively different state that features a slowly traveling passive background $\psi(x,t)$. At the same time, the LS in $ \phi(x,t)$ alternately oscillates \textit{and} travels at a faster speed into the opposite direction than the passive background. Similar to \cref{fig:sols_v0023_horz}(b), the (now traveling) LS almost fills the domain and the amplitude of the oscillations is much smaller at the center than at the edges of the LS. The LS travels by gaining density on one side and losing it on the other. 
The sites further away from the center of the LS again show alternately oscillating occupancy.

The various described patterns dominate in different regions of the snaking structure in \cref{fig:bd_v0023_horz}. Thereby, time simulations result in steady LSs in the central region in the range $-0.698 \lessapprox \bar\psi  \lessapprox -0.581$, while closer toward the ends of the snaking structure oscillatory states dominate. This can be seen in the magnifications shown as insets in \cref{fig:bd_v0023_horz}.

Results for a slightly higher activity $ v_0 = 0.3 $ are presented in Fig.~\ref{fig:bd_v003_horz}. For this value, all branches of steady LSs have vanished, i.e., we are in a region analogous to the one above the final cusp in Fig.~12 of Ref.\ \cite{HollAGKOT2021}. The periodic steady state is linearly stable at low densities as before. It becomes unstable via a Hopf bifurcation at $ \bar{\psi}\approx -0.717 $. Furthermore, at $ \bar{\psi}\approx-0.512 $ a branch of traveling periodic states emerges in a supercritical drift pitchfork bifurcation. It is initially unstable but eventually stabilized through a series of Hopf bifurcations with the final one occurring at $ \bar{\psi}\approx -0.434 $. Employing numerical path continuation of steady states, here, we only access resting and steadily traveling states. In the $ \bar{\psi}$-range where all of these states are unstable, we determine the system behavior by direct time simulations. As a result, we find five types of standing, traveling, and modulated traveling wave states, exemplified in the space-time plots in the insets of \cref{fig:bd_v003_horz}.

At high densities, two kinds of traveling states exist and coexist: Inset \ref{fig:bd_v003_horz}(a) shows a steadily traveling periodic state, also obtained by continuation. The second kind, presented in inset \ref{fig:bd_v003_horz}(b), consists of oscillating direction-reversing periodic states. The wiggling back and forth motion of density peaks is overlaid by a slow drift of the entire pattern. This is akin to the oscillating direction-reversing LSs found in Ref.\ \cite{OphausKGT2020} (see their Fig.~11(a)), albeit there they do not show an additional drift. Most likely, the corresponding branch emerges from the branch of steadily drifting states via a Hopf bifurcation. Note that there is a range of multistability.
Another type of wiggling state is presented in inset \ref{fig:bd_v003_horz}(c). There, two patches of wiggling peaks (having three peaks each and wiggling in anti-phase) are separated at the center and at the (periodic) boundary by a respective localized oscillating one-peak state. Thereby, the two one-peak states communicate via the wiggling patches such that together they form an alternating space-time pattern, similar to neighboring peaks in Fig.~\ref{fig:bd_v0023_horz}.
At first sight, the space-time plot in inset \ref{fig:bd_v003_horz}(d) seems to show a similar state. However, there, the fast interrelated oscillating and wiggling pattern is furthermore modulated by a slower oscillation, i.e., there is a second, lower frequency, i.e., a Hopf bifurcation is the most likely transition scenario. Finally, inset \ref{fig:bd_v003_horz}(e) represents a fully  space-time periodic checkerboard pattern where each peak oscillates and neighbors alternate in their oscillation.
 
Note that for all time-periodic states shown in the insets \ref{fig:bd_v003_horz}(a)-(d), the density of the passive particles $ \phi $ normally closely follows the one of the active species $ \psi $. An exception are nearly static density peaks in $ \phi $ whose sites are co-occupied by the oscillating peaks in $ \psi $ in insets \ref{fig:bd_v003_horz}(c)-(e).

Next, in \cref{fig:bd_v004_horz} we present the results at the highest here considered activity, $v_0=0.4$. As in \cref{fig:bd_v003_horz}, branches of resting and traveling periodic states are obtained by numerical continuation. The resting states are linearly stable at low densities $\bar\psi$, and become unstable via a Hopf bifurcation at $\bar\psi\approx-0.656$. Twelve further Hopf bifurcations occur along the branch (not shown). Then, at $\bar\psi\approx-0.486$ a branch of unstable steadily traveling periodic states emerges in a subcritical drift pitchfork bifurcation. In contrast to the case of \cref{fig:bd_v003_horz}, this branch remains unstable. Again, this leaves a large $\bar\psi$-range where all steady states are unstable and we resort to time simulations. At low densities, the time simulations produce the expected steady states. Directly beyond the first Hopf bifurcation, alternating periodic states are found similar to the ones presented in \cref{fig:bd_v003_horz}(e). Above $\bar\psi=-0.6$, two kinds of traveling states are found not known from the cases of lower activity. Both have in common that the passive background performs a wiggling back-forth motion, overlaid by a slow drift of the whole pattern, similar to Fig.\ \ref{fig:bd_v003_horz}(b). The active particles, however, move at a much larger speed in the same direction. The wiggling motion of the background is then due to a time-periodic pull the active particles ($\psi$) exercise onto the passive particles ($\phi$). At densities between $\bar\psi\approx-0.6$ and $\bar\psi\approx-0.44$ the active particles ($\psi$) organize into localized density peaks. Inset \ref{fig:bd_v004_horz}(a) shows such a state with three localized density peaks of $\psi$ (in red) that repeatedly move through the domain, and the background of passive particles ($\phi$, in blue) moving much slower. The number of peaks in $\psi$ increases with $\bar\psi$, until the whole domain is filled with density peaks. An example of such a traveling periodic state with different speeds of the active and passive particles is presented in inset \ref{fig:bd_v004_horz}(b). For both, localized and periodic traveling states with two speeds, the speeds increase with increasing mean density $\bar\psi$.

Finally, in \cref{fig:bd_v003_diag} we return to the lower activity of $v_0=0.3$ and analyze states occurring along the line through the phase diagram (\cref{fig:phasediagram}) defined by $\bar\psi = \bar\phi$. Contrary to the previous case where we had fixed $\bar\phi=0$, now the passive species no longer provides a periodic background. Instead, it shows similar density profiles as the active species.
At low densities $\bar\psi = \bar\phi$, the uniform liquid state is linearly stable. It is destabilized in a supercritical pitchfork bifurcation at $\bar\psi=\bar\phi\approx-0.721$, where the branch of resting periodic states with $n=8$ peaks emerges. Close to onset, these states are linearly stable, but very soon loose stability in a secondary pitchfork bifurcation at $\bar\psi=\bar\phi\approx-0.72$. There, two branches of symmetric resting LSs emerge subcritically, with an odd and even number of peaks, respectively. Both branches are initially unstable. The branch of odd LSs gains stability in a saddle-node bifurcation where it folds back toward higher densities. The branch of even LSs gains stability after a saddle node bifurcation and a pitchfork bifurcation, where the first of the ladder branches of asymmetric LSs emerges. Similar to the passive limit, the two branches of symmetric LSs undergo slanted snaking involving a series of saddle-node and pitchfork bifurcations. The latter produce in total five branches of asymmetric LSs. The branches of resting symmetric LSs reconnect to the branch of resting periodic states at $\bar\psi=\bar\phi\approx-0.486 $ when the whole domain is filled with peaks.  
In \cref{fig:bd_v0023_horz,fig:bd_v003_horz,fig:bd_v004_horz}, all LSs consist of crystalline patches of the active species ($\psi$) on a periodic background of the passive species ($\phi$). Here, all LSs are crystalline patches of an alloy of the passive and active species in a liquid background.

In contrast to the passive limit, at the chosen activity only the sub-branches of symmetric LSs with one, two, and three peaks are linearly stable, as further up each branch undergoes more than 30 Hopf bifurcations. The highest density for which resting LSs exist is $\bar\psi=\bar\phi\approx-0.741$.
Also the steady periodic states undergo further bifurcations: at $\bar\psi=\bar\phi\approx-0.598$ a branch of steadily traveling periodic states emerges in a supercritical drift pitchfork bifurcation. Initially it is unstable, but after undergoing a total of ten Hopf bifurcations it gains stability in a final one at $\bar\psi=\bar\phi\approx-0.485$. There is also a period-doubling pitchfork bifurcation on the branch of steadily traveling periodic states, where a branch of drifting states emerges, that are never stable and not shown here.

As before, this leaves a density range, where no resting or steadily traveling states exist.  Resorting to time simulations, at low densities up to $\bar\psi=\bar\phi=-0.76$ we recover first steady periodic states and then steady odd LSs. At higher densities, there are alternating LSs (not shown here), similar to those in \cref{fig:bd_v0023_horz,fig:sols_v0023_horz}, albeit on a liquid background. At a slightly higher mean density, inset \ref{fig:bd_v003_diag}(a) shows a state combining two localized patches of different states encountered before: At the center of the domain is a single wiggling peak performing a back-forth motion without net drift, while on the (periodic) boundary a single peak of the active species oscillates, i.e., behaves like a standing wave. Similar to the behavior of their domain filling variants, seen in \cref{fig:bd_v003_horz}, the passive density peak at the center closely follows the wiggling motion, while the passive density peak on the boundary oscillates with a very low amplitude only.

Another combination of two localized patches of alternating and wiggling LSs is presented in inset \ref{fig:bd_v003_diag}(c). There, a patch of two wiggling peaks coexists with a patch of an alternating LS with three peaks. This is furthermore overlaid by a very slow drift of the entire pattern. Note that the two outer peaks of the alternating LS are not at an equal distance to the wiggling LS. Because of this, both patches of LS are asymmetric, in contrast to those in inset \ref{fig:bd_v003_diag}(a) that show a distinct spatio-temporal symmetry. 
Inset \ref{fig:bd_v003_diag}(b) shows an example of the irregular, potentially chaotic states shown as star symbols in the main panel in \cref{fig:bd_v003_diag}. Here, we do not analyze these states further. 
Lastly, inset \ref{fig:bd_v003_diag}(d) gives a steadily traveling state consisting of a combination of two different LSs, with two and four peaks, respectively. The whole pattern moves at a constant speed with the passive density closely following the active one. Similar to the passive limit, at high densities the entire domain is filled with density peaks. There, only steadily traveling periodic states are found.

\section{\label{conc}Conclusions}
We have discussed a hierarchy of active PFC models that represent continuum descriptions for a mixture of active and passive colloidal particles at different levels of approximation. First, we have presented a systematic microscopic derivation of a very general PFC model (model~3a) from a DDFT, i.e., a microscopic continuum description that itself may be derived from a microscopic particle-based description \cite{WittkowskiL2011}. The derivation we have presented here includes a systematic treatment of the relevant orientational degrees of freedom. Thereby, our particular interest has been on the establishment of the nonlinear and coupling terms. Then, we have employed a series of approximations that have a clear physical meaning and motivation to simplify the general PFC model. Passing through these approximation steps, a hierarchy of models (models 3a-1c) has been established that allows for interesting insights into the microscopic justification of constructions used in various PFC models for active particles and mixtures, the approximations required for obtaining them, and possible generalizations. The minimal model (model~1) indeed corresponds to the one constructed and analyzed in Ref.\ \cite{HollAGKOT2021} by purely phenomenological means.

Note that, in passing, the presented derivation has also recovered a number of related models and contributed to the understanding of the employed approximations and their limitations. Namely, when eliminating the passive species from model~1 one obtains the active PFC model for a single species of active  colloidal particles derived and analyzed in Refs.\ \cite{MenzelL2013,MenzelOL2014,ChervanyovGT2016,OphausGT2018,OphausKGT2020,OphausKGT2020b}. Furthermore, a PFC model for a mixture of passive particles with a linear coupling between the two fields \cite{HollAT2020} (also see Section~4.1 in Ref.\ \cite{HollAGKOT2021}) is recovered when taking model~1 in the limit of vanishing activity. Note that this has allowed us to use the phase diagram obtained in Ref.\ \cite{HollAT2020} as a reference for the analysis of the parameter space. This implies that the various models 2 and 3 may in the passive limit be used as more exact models for mixtures of passive particles, the most precise being model 3a with $v_0=0$. (Note, however, that a coupling to $\vec{P}$ should not be present in the passive limit for spherical particles, see \cref{derivation2}.) In this way one may, e.g., obtain a model closely related to the one derived in Ref.\ \cite{TahaDMEH2019}. There, however, four-point correlations were not included. In contrast, the present model~3a contains entropic \textit{and} enthalpic fourth-order terms in the free energy functional.

In the second part of this work, we have restricted our attention to the derived minimal model~1. Linear stability analysis of the trivial uniform state at different mean concentrations has shown that it may become unstable to either a monotonic or an oscillatory small-scale instability similar to the one-component active PFC model \cite{OphausGT2018}. The difference is that now there can be two instability modes at similar or different wavenumbers active at the same time. These may be two monotonic modes or a monotonic and an oscillatory mode.

Beside the linear considerations, we have employed numerical continuation and time simulation methods to investigate the fully nonlinear regime. Thereby, the use of the former has allowed us to determine branches of periodic and localized steady and steadily traveling states. These can form the intricate intertwined slanted snaking structures expected for Swift-Hohenberg-type systems \cite{BurkeK2006,Knobloch2015} with conservation laws \cite{ThieleARGK2013,Knobloch2016,HollAGKOT2021}. However, in contrast to classical (conserved) Swift-Hohenberg-type systems, that are normally variational and therefore do only allow for steady states, here activity supports several types of standing, traveling, and modulated periodic and localized wave patterns. For instance, we have described direction reversing traveling periodic and localized states that we have called ``wiggling states'' as in Ref.\ \cite{OphausKGT2020} (where also other names and references for these states are given). Another state are spatio-temporal patterns where individual peaks behave like a standing wave and neighboring peaks show alternating (or anti-phase) oscillations. This resembles the modulated standing waves described in Ref.\ \cite{GolovinMBN1999} and has to our knowledge not yet been described for PFC systems. Here we have encountered such states in spatially periodic and localized versions. A recent study of nonreciprocally coupled Cahn-Hilliard equations reports on localized states whose outer peaks oscillate asymmetrically or symmetrically either in phase or in anti-phase \cite{FrohoffT2021}. Furthermore, the different regular spatio-temporal patterns can coexist in different regions of space, thereby giving rise to more intricate behavior. Transitions to seemingly irregular behavior have also been observed. 

A systematic investigation of the transitions between the various time-dependent states has been outside the scope of the present work, but forms a formidable future challenge.
Further possible extensions of this work include detailed comparative investigations of the more general models. An important remaining question is whether the minimal model~1 is already able to describe all states occurring in model 3a including their stability and sequence of appearance. Moreover, one could extend the derivation by incorporating also the nematic order parameter (as done in previous derivations for passive models \cite{Loewen2010,WittkowskiLB2010,WittkowskiLB2011,WittkowskiLB2011b}), particle inertia (as done for a single-species model in Refs.~\cite{AroldS2020,teVrugtJW2021}), or by considering a mixture of two different active species.

\section*{Data availability}
The raw data corresponding to the figures shown in this article are available as Supplementary Material \cite{SI}. 
\section*{Conflicts of interest}
There are no conflicts of interest to declare.

\begin{acknowledgments}
We thank Andrew J.\ Archer, Jens Bickmann, and Hartmut L\"owen for helpful discussions. M.t.V. thanks the Studienstiftung des deutschen Volkes for financial support. R.W.\ is funded by the Deutsche Forschungsgemeinschaft (DFG, German Research Foundation) -- WI 4170/3-1.
\end{acknowledgments}

\appendix
\section{\label{gradientexpansion}Gradient expansion}
Here, we briefly explain the method of gradient expansions \cite{YangFG1976}. We have to deal with terms of the form
\begin{equation}
\INT{}{}{^2r}\INT{}{}{^2r_1}g(\vec{r})h(\vec{r}_1)f(R),  
\end{equation}
where $g,h \in \{\psi,\phi,P_x,P_y\}$ and $R=\norm{\vec{r}-\vec{r}_1}$ with the Euclidean norm $\norm{\cdot}$. This is done in the standard way
\begin{align}
\nonumber &\INT{}{}{^2r}\INT{}{}{^2r_1}g(\vec{r})h(\vec{r}_1)f(R)\\ 
\nonumber &=\frac{1}{2\pi}\INT{}{}{^2r}\INT{}{}{^2r_1}\INT{}{}{^2k}g(\vec{r})h(\vec{r}_1)\hat{f}(\vec{k})e^{\ii\vec{k}\cdot(\vec{r}-\vec{r}_1)}\\
\nonumber &\approx\frac{1}{4\pi^2}\INT{}{}{^2r}\INT{}{}{^2r_1}\INT{}{}{^2k}g(\vec{r})h(\vec{r}_1)\\
&\quad\; (\hat{f}_0 -\hat{f}_2 k^2 + \hat{f}_4 k^4)e^{\ii\vec{k}\cdot(\vec{r}-\vec{r}_1)}\label{standardway}\\
\nonumber &=\frac{1}{2\pi}\INT{}{}{^2r}\INT{}{}{^2k}g(\vec{r})\hat{h}(\vec{k})(\hat{f}_0 - \hat{f}_2 k^2 + \hat{f}_4 k^4)e^{\ii\vec{k}\cdot\vec{r}}\\
\nonumber &=\INT{}{}{^2r} g(\vec{r})(\hat{f}_0 + \hat{f}_2\Nabla^2 + \hat{f}_4 \Nabla^4)h(\vec{r})
\end{align}
where
\begin{equation}
\hat{f}(\vec{k})=\frac{1}{2\pi}\INT{}{}{^2r}f(\norm{\vec{r}})e^{-\ii\vec{k}\cdot\vec{r}}.    
\end{equation}
After the $\approx$ sign in Eq.\ \eqref{standardway}, we have Taylor expanded $\hat{f}(\vec{k})$ up to fourth order in $\vec{k}$ about $\vec{k}=\vec{0}$, writing the second-order term with a minus sign and absorbing a factor $2\pi$ into the expansion coefficients for later convenience, and taking into account the symmetries arising from the fact that $f$ depends only on $R$. Note that for terms involving $\vec{P}$, only the zeroth-order contribution is considered. The expansion coefficients are given by
\begin{align}
\hat{f}_0 &=2\pi\INT{0}{\infty}{R}Rf(R),\\
\hat{f}_2&=\frac{\pi}{2}\INT{0}{\infty}{R}R^3 f(R),\\
\hat{f}_4&=\frac{\pi}{32}\INT{0}{\infty}{R}R^5 f(R).
\end{align}

\section{\label{coefficients}Expansion coefficients in Eq.\ (\ref{localform})}
The expansion coefficients in \cref{localform} are given by
\begin{align}
A_1 &= 8\pi^3\beta^{-1} \bar{\rho}_\psi^2\INT{0}{\infty}{R}R c_{\psi\psi}^{0}(R),\label{a1}\\
A_2 &= 2\pi^3\beta^{-1} \bar{\rho}_\psi^2 \INT{0}{\infty}{R}R^3 c_{\psi\psi}^{0}(R),\label{a2}\\
A_3 &= \frac{\pi^3}{8}\beta^{-1} \bar{\rho}_\psi^2 \INT{0}{\infty}{R}R^5 c_{\psi\psi}^{0}(R),\label{a3}\\
A_4 &= 2\pi^3\beta^{-1} \bar{\rho}_\psi^2\INT{0}{\infty}{R}R (c_{\psi\psi}^{1}(R) + c_{\psi\psi}^{-1}(R)),\label{a4}\\
A_5 &= 8\pi^2 \beta^{-1} \bar{\rho}_\psi \bar{\rho}_\phi\INT{0}{\infty}{R}R c_{\psi\phi}(R),\label{a5}\\
A_6 &= 2\pi^2 \beta^{-1} \bar{\rho}_\psi \bar{\rho}_\phi\INT{0}{\infty}{R}R^3 c_{\psi\phi}(R),\label{a6}\\
A_7 &= \frac{\pi^2}{8} \beta^{-1} \bar{\rho}_\psi \bar{\rho}_\phi\INT{0}{\infty}{R}R^5 c_{\psi\phi}(R),\label{a7}\\
A_8 &= 2\pi\beta^{-1}\bar{\rho}_\phi^2\INT{0}{\infty}{R}R c_{\phi\phi}(R),\label{a8}\\
A_9 &= \frac{\pi}{2}\beta^{-1}\bar{\rho}_\phi^2 \INT{0}{\infty}{R}R^3 c_{\phi\phi}(R),\label{a9}\\
A_{10} &= \frac{\pi}{32}\beta^{-1}\bar{\rho}_\phi^2 \INT{0}{\infty}{R}R^5 c_{\phi\phi}(R),\label{a10}\\
A_{11} &= 16\pi^5\beta^{-1} \bar{\rho}_\psi^3\INT{0}{\infty}{R}\INT{0}{\infty}{R'} R R' c_{\psi\psi\psi}^{0,0}(R,R'),\label{a11}\\
A_{12} &= 4\pi^5\beta^{-1} \bar{\rho}_\psi^3\INT{0}{\infty}{R}\INT{0}{\infty}{R'} R R' \notag \\ &\quad\,(c_{\psi\psi\psi}^{0,1}(R,R')+c_{\psi\psi\psi}^{0,-1}(R,R') \label{a12}\\ &\quad\,+c_{\psi\psi\psi}^{1,0}(R,R')+c_{\psi\psi\psi}^{-1,0}(R,R')\notag\\
&\quad\,+2c_{\psi\psi\psi}^{1,-1}(R,R')),\notag\\ 
A_{13} &= 24\pi^4 \beta^{-1} \bar{\rho}_\psi^2\bar{\rho}_\phi\INT{0}{\infty}{R}\INT{0}{\infty}{R'}R R' c_{\psi\psi\phi}^{0}(R,R'),\label{a13}\\
A_{14} &= 6\pi^4\beta^{-1} \bar{\rho}_\psi^2\bar{\rho}_\phi \INT{0}{\infty}{R}\INT{0}{\infty}{R'}R R' \notag \\
&\quad\,(c_{\psi\psi\phi}^{1}(R,R') + c_{\psi\psi\phi}^{-1}(R,R')),\label{a14}\\
A_{15} &= 12\pi^3 \beta^{-1}\bar{\rho}_\psi\bar{\rho}_\phi^2 \INT{0}{\infty}{R}\INT{0}{\infty}{R'}R R' c^{0}_{\psi\phi\phi}(R,R'),\label{a15}\\
A_{16} &= 2\pi^2 \beta^{-1} \bar{\rho}_\phi^3 \INT{0}{\infty}{R}\INT{0}{\infty}{R'}R R' c_{\phi\phi\phi}(R,R'),\label{a16}\\
A_{17} &= \frac{64}{3}\pi^7\beta^{-1}\bar{\rho}_\psi^4\INT{0}{\infty}{R}\INT{0}{\infty}{R'}\INT{0}{\infty}{R''} R R'R'' \notag \\ &\quad\, c^{0,0,0}_{\psi\psi\psi\psi}(R,R',R''),\label{a17}\\
A_{18} &= \frac{16}{3}\pi^7\beta^{-1}\bar{\rho}_\psi^4\INT{0}{\infty}{R}\INT{0}{\infty}{R'}\INT{0}{\infty}{R''} R R'R'' \notag\\ 
&\quad\, (c^{0,0,1}_{\psi\psi\psi\psi}(R,R',R'')+c^{0,0,-1}_{\psi\psi\psi\psi}(R,R',R'') \notag\\
&\quad\, +c^{0,1,0}_{\psi\psi\psi\psi}(R,R',R'')+c^{0,-1,0}_{\psi\psi\psi\psi}(R,R',R'') \label{a18}\\
&\quad\,+c^{-1,0,0}_{\psi\psi\psi\psi}(R,R',R'') +c^{1,0,0}_{\psi\psi\psi\psi}(R,R',R'') \notag\\ 
&\quad\, +2c^{1,-1,0}_{\psi\psi\psi\psi}(R,R',R'')+2 c^{0,-1,1}_{\psi\psi\psi\psi}(R,R',R'')\notag \\ &\quad\,+2c^{-1,0,1}_{\psi\psi\psi\psi}(R,R',R'')),\notag\\
A_{19} &= 4\pi^7\beta^{-1}\bar{\rho}_\psi^4\INT{0}{\infty}{R}\INT{0}{\infty}{R'}\INT{0}{\infty}{R''} R R'R'' \notag\\& \quad\, (c^{1,1,1}_{\psi\psi\psi\psi}(R,R',R'')+c^{-1,1,1}_{\psi\psi\psi\psi}(R,R',R'')),\label{a19}\\
A_{20} &= \frac{128}{3}\pi^6\beta^{-1} \bar{\rho}_\psi^3\bar{\rho}_\phi\INT{0}{\infty}{R}\INT{0}{\infty}{R'}\INT{0}{\infty}{R''} R R' R'' \notag\\
&\quad\, c_{\psi\psi\psi\phi}^{0,0}(R,R',R''),\label{a20}\\
A_{21} &= \frac{32}{3}\pi^6\beta^{-1} \bar{\rho}_\psi^3\bar{\rho}_\phi\INT{0}{\infty}{R}\INT{0}{\infty}{R'} \INT{0}{\infty}{R''} R R' R'' \notag \\
&\quad\,(c_{\psi\psi\psi\phi}^{0,1}(R,R',R'')+c_{\psi\psi\psi\phi}^{0,-1}(R,R',R'') \label{a21}\\ &\quad\,+c_{\psi\psi\psi\phi}^{1,0}(R,R',R'')+c_{\psi\psi\psi\phi}^{-1,0}(R,R',R'')\notag\\
&\quad\,+2c_{\psi\psi\psi\phi}^{1,-1}(R,R',R'')),\notag\\ 
A_{22} &= 32\pi^5 \beta^{-1} \bar{\rho}_\psi^2 \bar{\rho}_\phi^2 \INT{0}{\infty}{R}\INT{0}{\infty}{R'}\INT{0}{\infty}{R''} R R' R''\notag\\
&\quad c_{\psi\psi\phi\phi}^{0}(R,R',R''),\label{a22}\\
A_{23} &= 8\pi^5 \beta^{-1} \bar{\rho}_\psi^2 \bar{\rho}_\phi^2 \INT{0}{\infty}{R}\INT{0}{\infty}{R'}\INT{0}{\infty}{R''} R R' R'' \notag\\
&\quad\, (c_{\psi\psi\phi\phi}^{1}(R,R',R'') +c_{\psi\psi\phi\phi}^{-1}(R,R',R'')),\label{a23}\\
\begin{split}
A_{24}&= \frac{32}{3}\pi^4\beta^{-1} \bar{\rho}_\psi\bar{\rho}_\phi^3\INT{0}{\infty}{R}\INT{0}{\infty}{R'}\INT{0}{\infty}{R''} R R' R'' \\
&\quad\, c_{\psi\phi\phi\phi}(R,R',R''),
\label{a24}\end{split}\\
\begin{split}
A_{25}&= \frac{4}{3}\pi^3 \beta^{-1}\bar{\rho}_\phi^4\INT{0}{\infty}{R}\INT{0}{\infty}{R'}\INT{0}{\infty}{R''} R R' R'' \\
&\quad\, c_{\phi\phi\phi\phi}(R,R',R'').
\label{a25}\end{split}
\end{align}
To simplify \cref{a12,a18,a19,a21}, we have used the fact that replacing $m_i'$ by $-m_i'$ in \cref{fouriercoefficients} leads to a complex conjugation.

\section{\label{coefficients2}Expansion coefficients in Eq.\ (\ref{largefreeenergy1})}
The expansion coefficients in \cref{largefreeenergy1} are given by
\begin{align}
\tilde{A}_3 &= - A_3,\label{tildea3}\\
\nonumber \tilde{A}_4 &= \frac{B_1 - 2A_4}{2}- \frac{2}{3}(A_{14}\Delta\phi + A_{12}\Delta\psi)\\
&\EinrueckabstandI-\frac{1}{2}(A_{23}(\Delta\phi)^2 + B_1\Delta\psi+A_{21}\Delta\phi\Delta\psi \\ 
\nonumber &\EinrueckabstandI+ (A_{18} - B_1) (\Delta\psi)^2),\\
\tilde{A}_7 &= - A_7,\\
\tilde{A}_{10}&= - A_{10},\\
\tilde{A}_{12}&= A_{12} + \frac{3}{2}A_{18}\Delta\psi + \frac{3}{4}A_{21}\Delta\phi+ \frac{3}{4}B_1,\\
\tilde{A}_{13}&= A_{13} + \frac{9}{4}A_{20}\Delta\psi + \frac{3}{2}A_{22}\Delta\phi,\\
\tilde{A}_{14}&= A_{14} + \frac{3}{4}A_{21}\Delta\psi + \frac{3}{2}A_{23}\Delta\phi,\\
\tilde{A}_{15}&= A_{15} + \frac{9}{4}A_{24}\Delta\phi+ \frac{3}{2}A_{22}\Delta\psi,\\
\tilde{A}_{17}&= \frac{B_1}{3}-A_{17},\\
\tilde{A}_{18}&= A_{18} - B_1,\\
\tilde{A}_{19} &= \frac{B_1 - 8 A_{19}}{8},\\
\tilde{A}_{20}&= A_{20},\\
\tilde{A}_{21}&= A_{21},\\
\tilde{A}_{22}&= A_{22},\\
\tilde{A}_{23}&= A_{23},\\
\tilde{A}_{24}&= A_{24},\\
\tilde{A}_{25}&= \frac{B_2}{3}-A_{25},\\
\tilde{\epsilon}_\psi &= B_1 - A_1 - A_3\tilde{q}_\psi^4 - (B_1 + 2A_{11})\Delta \psi \notag\\
&\quad\, + (B_1 - 3A_{17})(\Delta\psi)^2  - \frac{2}{3}A_{13}\Delta\phi\label{epspsi}\\
&\quad\,-\frac{3}{2}A_{20}\Delta\psi\Delta\phi-\frac{1}{2}A_{22}(\Delta\phi)^2,\notag\\
\tilde{\epsilon}_\phi &= B_2 - A_8 - A_{10}\tilde{q}_\phi^4 - (B_2 + 2A_{16})\Delta \phi  \notag\\
&\quad\, + (B_2 - 3A_{25})(\Delta\phi)^2 -\frac{2}{3}A_{15}\Delta\psi \label{epsphi}\\
&\quad\,-\frac{3}{2}A_{24}\Delta\psi\Delta\phi-\frac{1}{2}A_{22}(\Delta\psi)^2,\notag\\
\nonumber \tilde{\epsilon}_{\mathrm{coup}} &= -\frac{1}{2}A_5 -\frac{2}{3}(A_{13}\Delta\psi + A_{15}\Delta\phi)\\
&\quad\,-\frac{3}{4}(A_{20}(\Delta\psi)^2 + A_{24}(\Delta\phi^2))\label{tildeepsiloncoup}\\
&\quad\, - A_{22}\Delta\psi\Delta\phi -A_7 \tilde{q}_{\mathrm{coup}}^4,\notag\\
\tilde{q}_\psi &= \sqrt{-\frac{A_2}{2 A_3}},\\
\tilde{q}_\phi &= \sqrt{-\frac{A_9}{2 A_{10}}},\\
\tilde{q}_{\mathrm{coup}} &= \sqrt{-\frac{A_6}{2A_7}}.
\end{align}
Note that the numbering of the $\tilde{A}_i$ does not contain all numbers between 1 and 25. For example, there is no coefficient $\tilde{A}_1$ or $\tilde{A}_2$. The reason for this is that we wish the numbering of the $\tilde{A}_i$ to be consistent with the numbering of the $A_i$ in \cref{coefficients}. For example, the coefficient defined in terms of $A_3$ is called $\tilde{A}_3$ and not $\tilde{A}_1$ (cf. \cref{tildea3}).

\section{\label{rescaling}Nondimensionalized coefficients}
The coefficients of model 3a (\cref{freeenergymodel3a,psi4,p4,phi4}) are given by
\begin{align}
\epsilon_\psi &= \beta r_0^2 \psi_0^2\tilde{\epsilon}_\psi ,\\
\epsilon_\phi &= \beta r_0^2 \phi_0^2\tilde{\epsilon}_\phi ,\\
\epsilon_{\mathrm{coup}} &= \beta r_0^2 \psi_0 \phi_0\tilde{\epsilon}_{\mathrm{coup}},\\
q_\psi &= r_0\tilde{q}_\psi,\\
q_\phi &= r_0 \tilde{q}_\phi,\\
q_{\mathrm{coup}} &= r_0\tilde{q}_{\mathrm{coup}},\\
v_0 &= \frac{v t_0}{\sqrt{2} r_0},\\
C_1 &= \beta r_0^2 P_0^2\tilde{A}_4,\\ 
C_2 &= \beta  r_0^2 P_0^4\tilde{A}_{19},\\
C_3 &= -\frac{1}{3}\beta r_0^2 \psi_0 P_0^2\tilde{A}_{12}, \\
C_4 &= -\frac{1}{4}\beta r_0^2 \psi_0^2 P_0^2\tilde{A}_{18}, \\
C_5 &=  -\frac{1}{3}\beta r_0^2 \phi_0 P_0^2\tilde{A}_{14},\\
C_6 &= -\frac{1}{4}\beta r_0^2 \psi_0^2 P_0^2A_{23}, \\
C_7 &= -\frac{1}{4}\beta r_0^2 \psi_0 P_0^2 \phi_0 A_{21}, \\
G_\phi&= \frac{\tilde{A}_{10}}{\tilde{A}_3}\sqrt{\frac{\tilde{A}_{17}}{\tilde{A}_{25}}}\label{lphi},\\
G_{\mathrm{coup}}&=\frac{\tilde{A}_7 }{\tilde{A}_3}\sqrt[4]{\frac{\tilde{A}_{17}}{\tilde{A}_{25}}},\\
D_r &= \frac{r_0^2 D_R}{D_0},\\
M_\phi &= \frac{2\pi\bar{\rho}_\psi D}{\bar{\rho}_\phi D_0}\sqrt{\frac{\tilde{A}_{25}}{\tilde{A}_{17}}},\label{mphi}\\ 
a_2 &= -\frac{1}{3}\beta r_0^2 \psi_0^2 \phi_0\tilde{A}_{13},\\
a_3 &= -\frac{1}{3}\beta r_0^2 \psi_0 \phi_0^2\tilde{A}_{15},\\
a_4 &= -\frac{1}{4}\beta r_0^2 \psi_0^3 \phi_0A_{20}, \\
a_5 &= -\frac{1}{4}\beta r_0^2 \psi_0^2 \phi_0^2A_{22}, \\
a_6 &= -\frac{1}{4}\beta r_0^2 \psi_0 \phi_0^3 A_{24}.
\end{align}

\bibliography{refs}
\end{document}